\documentclass[bibyear]{aa}
\usepackage{graphicx}
\usepackage{txfonts}
\usepackage{natbib}
\bibpunct{(}{)}{;}{a}{}{,}
\usepackage{scrextend}			
\usepackage{tikz}
\usepackage{multirow}
\usepackage{hyperref}

\definecolor{Acol}{RGB}{7,32,244}
\definecolor{Bcol}{RGB}{255,0,23}
\definecolor{Ccol}{RGB}{0,127,36}
\definecolor{Dcol}{RGB}{0,192,191}
\definecolor{Ecol}{RGB}{196,14,184}

\begin{document} 

\newcommand{\Flux}{\ensuremath{\mathrm{F}}}
\newcommand{\ADU}{\ensuremath{\mathrm{ADU}}}
\newcommand{\Gain}{\ensuremath{\mathrm{G}}}
\newcommand{\ENF}{\ensuremath{\mathrm{ENF}}}
\newcommand{\be}{\ensuremath{\beta}}
\newcommand{\R}{\ensuremath{R_0}}
\newcommand{\chis}{\ensuremath{\chi^2_r}}
\newcommand{\comments}[1]{{\color{red} #1}} 
\newcommand{\tbd}{{\color{red} ...TBD...}} 
\newcommand\todo[1]{{\color{red} \bf #1}}
\newcommand\K[1]{{\color{blue} \bf #1}} 

\newcommand{\cyp}[1]{{\color{green} #1}}

\newcommand\NA[1]{{\color{pink} \bf #1}}

\title{Modeling the e-APD SAPHIRA/C-RED ONE camera at low flux level}
\subtitle{An attempt of photon counting in the near-infrared with the MIRC-X interferometric combiner}


\titlerunning{Modeling the SAPHIRA/C-RED ONE camera at low flux level}

   \author{C. Lanthermann\inst{1}
             \and
             N. Anugu\inst{2}
          \and
          J.-B. Le Bouquin\inst{1, 3}
          \and
          J.D. Monnier\inst{3}
          \and
          S. Kraus\inst{2}
          \and
          K. Perraut\inst{1}
          }

   \institute{Univ. Grenoble Alpes, CNRS, IPAG, 38000 Grenoble, France
     \and
     University of Exeter, School of Physics and Astronomy, Stocker Road, Exeter, EX4 4QL, UK
     \and
     University of Michigan,Department of Astronomy, 1085 S. University Ave; West Hall 323; Ann Arbor, MI 48109
\\ correspondence: \texttt{cyprien.lanthermann@univ-grenoble-alpes.fr}
   }

   \date{Received 10/01/2019; accepted 15/03/2019}

 
  \abstract
   {We implement an electron avalanche photodiode (e-APD) in the MIRC-X instrument, upgrade of the 6-telescope near-infrared imager MIRC, at the CHARA array. This technology should improve the sensitivity of near-infrared interferometry.}
   {We characterize a near-infrared C-RED ONE camera from First Light Imaging (FLI) using an e-APD from Leonardo (previously SELEX).}
   {We first used the classical Mean-Variance analysis to measure the system gain and the amplification gain. We then developed a physical model of the statistical distribution of the camera output signal. This model is based on multiple convolutions of the Poisson statistic, the intrinsic avalanche gain distribution, and the observed distribution of the background signal. At low flux level, this model constraints independently the incident illumination level, the total gain, and the excess noise factor of the amplification.}
   {We measure a total transmission of $48\pm3\%$ including the cold filter and the Quantum Efficiency.
   We measure a system gain of 0.49 ADU/e, a readout noise of $10$ ADU, and amplification gains as high as 200. These results are consistent between the two methods and therefore validate our modeling approach. The measured excess noise factor based on the  modeling is $1.47\pm0.03$, with no obvious dependency with flux level or amplification gain.}
   {The presented model allows measuring the characteristics of the e-APD array at low flux level independently of preexisting calibration. With $<0.3$\,electron equivalent readout noise at kilohertz frame rates, we confirm the revolutionary performances of the camera with respect to the PICNIC or HAWAII technologies. However, the measured excess noise factor is significantly higher than the one claimed in the literature ($<$1.25), and explains why counting multiple photons remains challenging with this camera.}


   \maketitle


\section{Introduction}
\label{sec:intro}

Optical long baseline interferometry allows imaging at higher angular resolution than classical monolithic telescopes for a given wavelength. For instance, the gain in resolution is about a factor 3 (resp., 8) between a diffraction-limited 40-m single-dish telescope and the Very Large Telescope Interferometer \citep[VLTI, ][]{2012SPIE.8445E..0DH} (resp., the CHARA array; \citet{2018SPIE10701E..02G}). As it is a natural trade-off between the damaging effect of atmospheric turbulence at shorter wavelengths and the increased sky brightness at longer wavelengths, many interferometric instruments operate in the near-infrared (NIR), like PIONIER and GRAVITY at the VLTI \citep{2011A&A...535A..67L,2017A&A...602A..94G}; CLIMB, CLASSIC, JouFLU, and MIRC-X at CHARA \citep{2006SPIE.6268E..1PM,2012SPIE.8445E..3CT,2014SPIE.9146E..1AS}. As a typical example, single-mode optical interferometry in the H-band (1.6~$\mu$m) requires individual exposure times of about 10~ms. In this spectral range, to cope with atmospheric effects and to stabilize the interferometric fringe patterns implies to measure and servo-control the fringe positions during a fraction of the coherence time, i.e.\ during less than a few milliseconds. Fringe tracking in the NIR thus needs fast detectors working at kHz speed and allows to push the instrument sensitivity by allowing minute-long exposures to be done on the scientific detector \citep{2017A&A...602A..94G}.

In addition, in the H-band, the total instrument and sky background is about 10 photons per exposure per telescope. Considering that at least $100 \times 10$ pixels are commonly needed to encode the spectrally dispersed fringe signal, the background noise is much lower than one photon per pixel and per exposure. Typical readout noise of PICNIC or HAWAII detectors for such frame rate and format are $\sim$8\,electrons per exposure and per pixel \citep{2004PASP..116..377P,2014SPIE.9146E..1WB}. Consequently, the detector readout noise has remained the limiting factor for decades.

The situation has changed in the 2000s with the development of matrices of electron Avalanche Photo-Diodes (e-APD) in the NIR. These detectors promised to deliver sub-electron readout noise thanks to amplification of the signal before the readout. On the one hand, a French collaboration of industrial (SOFRADIR, CEA-LETI) and academic (IPAG, LAM, ONERA) partners designed and built several prototypes of the so-called RAPID detector \citep{2014SPIE.9148E..18F,2014SPIE.9146E..1NG}. On the another hand, the Leonardo company (former SELEX) in collaboration with the ESO developed the SAPHIRA detectors and used as wavefront sensors and a fringe tracker for the GRAVITY instrument \citep{gert2016,2017A&A...602A..94G}. This detector was made available as a commercial product, called C-RED ONE, by the First Light Imaging company \citep{2016SPIE.9907E..2EG}. 

Within the framework of the installation of adaptive optics on the CHARA array \citep{CHARA} and with the aim of pushing the sensitivity of high-angular resolution imaging in the NIR, we purchased the first two C-RED ONE cameras in 2017. The first one has been used to upgrade the H-band 6-telescope combiner MIRC, now called MIRC-X \citep{MIRC, MIRCX}, at the focus of the CHARA optical interferometer since mid-2017. The second one will be used in the K-band copy of this instrument, called MYSTIC, and expected to be installed in 2019 at CHARA \citep{MYSTIC}.

The goal of this paper is to measure the characteristics of the e-APD detector of MIRC-X at our typical low flux level and to determine if photon-counting is achievable in realistic operational conditions.

After this introduction, the second section describes the C-RED ONE camera and the e-APD technology. In the third section, we present a classical characterization of the detector noise and amplification of the camera, based on the Mean-Variance method. In the fourth section, we introduce an innovative model of the signal distribution which allows us to extract the excess noise factor and the gain. The fifth section discusses the obtained results and compares them with the literature, especially the significant discrepancy of the excess noise factor. The paper ends with a brief summary and conclusion.

\section{The C-RED ONE camera}
\label{sec:setup}

\subsection{e-APD technology}\label{sec:apdtech}

The e-APD technology consists in applying a bias voltage in a lower layer of the pixel. The electron generated by the incident photon in the absorption region migrates to the bottom of the pixel. When the electron travels through the multiplication region, it is accelerated by the bias voltage. At some point, the kinetic energy is high enough to ionize an atom of the substrate by a collision. The two electrons are accelerated again, generating new electrons by collision. This avalanche process is explained in further details in \cite{gert2010} and \cite{gert2012}.

The avalanche process is a stochastic process. An incident photon can lead to different numbers of final electrons with a given probability distribution. In this paper, we call this probability distribution the \emph{avalanche gain distribution} $M$. From this amplification gain distribution, we can derive two principal characteristics of the system: the mean avalanche gain and the excess noise factor. The mean avalanche gain is defined by
\begin{equation}\label{eq:GM}
\Gain_{\mathrm{av}} = \langle M \rangle.
\end{equation}
The excess noise factor (ENF) is defined by 
\begin{equation}\label{eq:ENFM}
\ENF = \frac{\langle M ^{2}\rangle}{\langle M \rangle ^{2}}.
\end{equation}
The \ENF{} is the additional noise brought by the avalanche process on the output signal, compared to the photon noise obtained without the avalanche process. In the following, we also make use of the mean total gain, defined as:
\begin{equation}\label{eq:Gtotal}
\Gain = \Gain_{\mathrm{sys}}\,\Gain_{\mathrm{av}},
\end{equation}
where $\Gain_{\mathrm{sys}}$ is the system gain in $\ADU/\mathrm{e}$.

\subsection{Camera readout}

\begin{figure*}[!ht]
\centering
\includegraphics[scale=0.5]{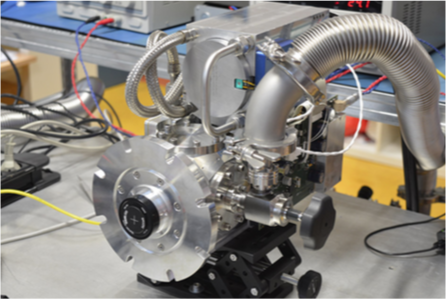}
\includegraphics[scale=0.25]{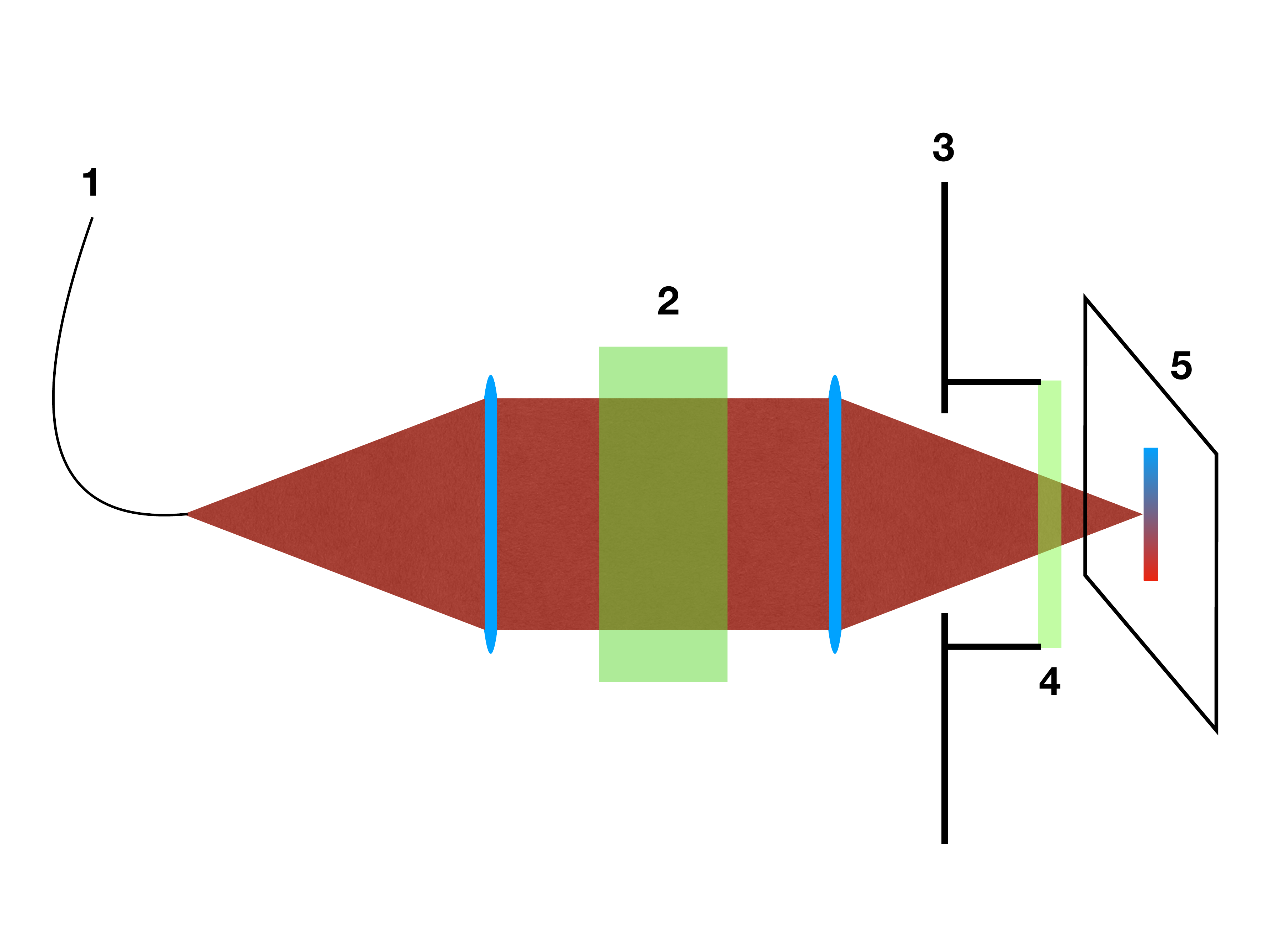}
\caption{{\bf Left.} Picture of the MIRC-X C-RED ONE. {\bf Right.} Scheme of the experimental setup. 1: H band single-mode fiber fed by a white-light source; 2: Dispersion prism; 3: Camera cold stop and baffle; 4: Low pass cold filter; 5: Detector.} \label{fig:scheme}
\end{figure*}

Our C-RED ONE cameras (Fig.~\ref{fig:scheme}-left) embark SAPHIRA MCT SWIR Mark13 e-APD 320$\times{}$256 pixels detectors whose overall characteristics are summarized in Tab.~\ref{tab:FLIspec}.

{\renewcommand{\arraystretch}{1.3}
\begin{table}[!ht]
\caption{Characteristics of the SAPHIRA MCT SWIR Mark13 detector embarked in our C-RED ONE camera, as specified by the manufacturer.}
\label{tab:FLIspec}
\begin{center}      
\begin{tabular}{p{55mm}p{25mm}} 
\hline \hline 
Characteristics & Values\\
\hline
  Format &  320 $\times$ 256  pixels\\
  Detector sensitivity & 0.8 to 3.5 $\mu$m  \\
  Readout Speed, full frame single read  & 3500 frames/s  \\
  Total noise for 1 ms of integration, looking at a 300 K scene & 0.4 e/pixel \\
  Quantum Efficiency & $>$ 70 $\%$ \\
  Cold filters transmission (at 1.55 $\mu$m) & 81 $\%$ \\
    H-band filter cutoff & 1.739 $\mu$m \\
  Excess Noise Factor (ENF) & 1.25 \\
  System Gain & 0.59 \ADU{}/e \\
  Operating temperature & 80 K \\
\hline
\end{tabular}
\end{center}
\end{table}
}

The camera offers various readout modes. Our custom readout is inspired from the PICNIC camera from the IOTA observatory \citep{PICNIC}. First, it consists of multiple consecutive reads of the same pixel (NREADS) before moving to the next pixel on the row. A row is then read multiple times (NLOOPS) before moving to the next row. Those multiple NREADS and NLOOPS are then averaged. It reduces the readout noise and extends the integration time while keeping a somewhat simultaneous integration window for the various pixels of the same row. In the following, we call a \emph{frame} the process of reading the entire array once, with this combination of NREADS and NLOOPS. Finally, the array is read multiple times non-destructively (NFRAMES\_PER\_RESET) before the detector is reset and the process starts again. Note that it is possible to read sub-windows of the full pixel array. Cropping is possible in the row and column directions and this increases the frame rate.

In most implementations of the on-the-ramp mode, the set of NFRAMES\_PER\_RESET along the ramp is collapsed into a single value of flux per pixel by the mean of a linear fit. This is not the case in interferometric instruments where a fast frame rate is necessary \citep{2004PASP..116..377P,2011A&A...535A..67L}. Instead, we analyze the flux of each frame individually in the ramp, by subtracting the measurement of the previous frame.

A typical setup used for on-sky observations consists of reading a window of 320~$\times$~17 pixels, with NREADS~=~12 and NLOOPS~=~8. It provides a frame rate of $\approx 355$~Hz. The detector is generally reset every 100 frames only. In the experiments of this paper, we used NREADS~=~8 and NLOOPS~=~2 in order to increase the frame rate to $\approx 1916$~Hz. This was necessary to lower the background signal per pixel and per frame. Each data set has a total of 2000 frames.

Before being analyzed, we correct the RAW data from a parasitic electronic signal from the pulse-tube, which appears as a nearly sinusoidal additive signal. The process is explained in~\cite{SPIE2018}. We then compute the flux per frame, in \ADU{}, by taking the difference between two consecutive frames.

\subsection{Cold optics}\label{sec:optic}

The C-RED ONE camera is delivered in a standalone cryostat, operating at 80~K thanks to a pulse-tube. It has an f/4 cold aperture at about 35\,mm from the detector and 15\,mm from the external surface of the cryostat window (e.g., where the last warm optics can be located). Our MIRC-X camera holds four low-pass cold filters to remove the thermal background. Two have a cut-off wavelength at 1.739\,$\mu$m, and two at 2.471\,$\mu$m. The latter is mandatory to increase the rejection at wavelengths longer than 3$\,\mu$m. The theoretical amount of electrons due to the thermal background flux seen by a pixel in a single frame can be computed as follows:
\begin{equation}
\Flux_{\mathrm{BKG_{thermal}}} = \int_0^\infty\,B(\lambda,300\mathrm{K})\;A\;\Omega\;T(\lambda)\;QE(\lambda)\;\frac{1}{f}\;\mathrm{d}\lambda,
\end{equation}https://www.overleaf.com/project/5b8e824862a9062358c47262
where $B(\lambda,300\mathrm{K})$ is the blackbody emission at 300\,K, $A~=~(24\,\mu\mathrm{m})^2$ is the surface of a pixel, $\Omega = \pi/64$~str is the solid angle corresponding to the cold aperture, $T(\lambda)$ is the combined transmission of the filters, $QE(\lambda)$ is the quantum efficiency, and $f=1916$\,Hz is the frame rate. 
We measured the combined transmission and quantum efficiency of the camera at 1.55 $\mu$m by the mean of a NIST-traceable calibrated diode. The result is $48\pm 3$\%. This value is somewhat in agreement with the theoretical specifications of 55\% at this wavelength.
Using the theoretical transmission and quantum efficiency values gives an incident background on the detector of  $\Flux_{\mathrm{BKG_{thermal}}}~=~0.042$\,e/frame/pixel. 

\subsection{Warm optics setup}

The data set used for the model presented in this paper have been taken with the optical configuration used for on-sky observations, but with the instrument fed by an internal thermal white-light source (Fig.~\ref{fig:scheme}-right). In summary, the light is injected into a fiber whose output acts as the entrance slit of a simple spectrograph. A prism disperses the light over 12\,pixels across the H-band. The image on the detector, therefore, consists of $3 \times 12$ illuminated pixels. The neighboring pixels are not illuminated by the flux coming from the fiber. In the following, we used only the value of a single pixel of the spectrum, which receives the flux from a sharp spectral band of approximately 0.05\,$\mu m$ centered on 1.55\,$\mu m$.


\section{Detector characterization}\label{sec:detectcharac}

\subsection{Mean-Variance analysis}
\label{an:ptc}

The Mean-Variance method is a very common way of measuring the gain and the readout noise of a detector \citep{PTC}. It uses the fact that the temporal variance of a photometrically stable signal is proportional to the flux signal. The proportionality coefficient is the total gain times the excess noise factor. The interception of the fit with the y-axis gives the readout noise. To measure the system gain of our detector, we perform a Mean-Variance curve at an avalanche gain of 1 (i.e., without avalanche process).

\begin{figure}[!ht]
\centering
\includegraphics[scale=0.48]{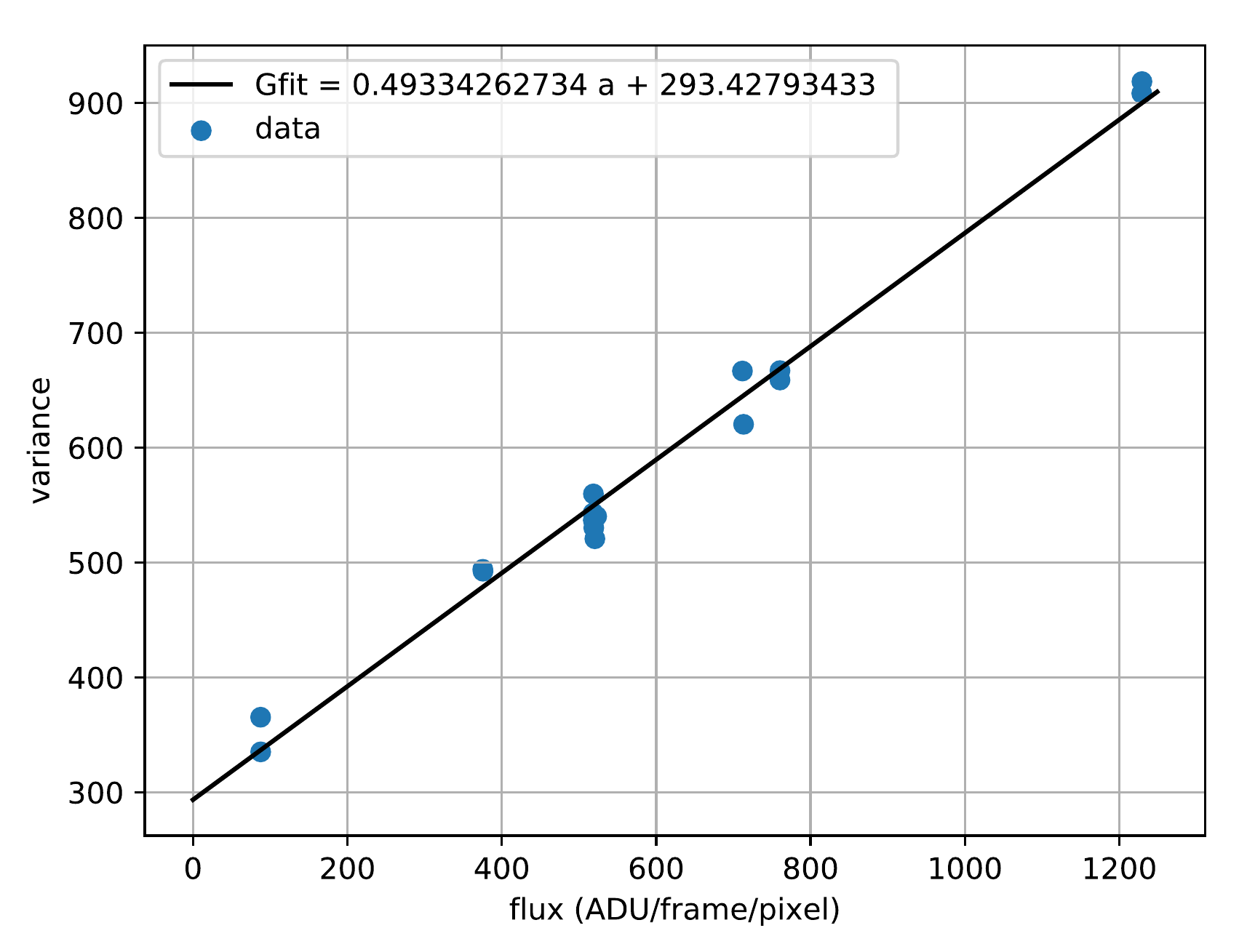}
\caption{Mean-Variance curve at avalanche gain = 1 (no avalanche process). The horizontal axis shows the mean flux in ADU/frame/pixel while the vertical axis gives the temporal variance. The blue dots are the data points. The black line is the fit of a linear function on the data points.\label{fig:PTC}}
\end{figure}

We measure a system gain $\Gain_{sys} = 0.49\,\ADU/\mathrm{e}$. Note that this is somewhat different from the expected system gain of $0.59\,\ADU/\mathrm{e}$ provided on the data sheet from FLI. We dig into the calibration data of the manufacturer. Depending on the portion of the curve that we fit, the gain can vary from 0.42 to 0.61.  These differences may be explained by non-linearity of the detector at different flux levels.  We do not study the effects of the non-linearity in this paper because we focus on the low-light-level regime.

The Mean-Variance analysis also allows estimating the detector readout noise, as the floor noise obtained without illumination. The readout noise is $\approx 17\,\ADU$ RMS, derived from the variance of 293 $\ADU{}^{2}$ from the fit. Using the system gain, this corresponds to $\approx 35\,\mathrm{e}$, which is comparable to the $45\,\mathrm{e}$ measured with no APD gain by \citet{gert2016} on a  similar device and with a similar method.

We have identified that this detector noise is in fact partially correlated between the pixels. When using proper filtering \citep{SPIE2018}, we typically achieve detector noise as low as $\approx 12\,\ADU$ RMS.

\subsection{Avalanche Gain}
\label{sec:avgaincal}

The avalanche gain is the mean value of electrons at the end of the avalanche process for one incident photo-electron. It represents the amplification of the incoming signal before the readout. To calibrate the avalanche gain, we measure the flux registered by the camera at different requested avalanche gains and divide it by the flux measured with no avalanche gain for the same incident flux. At the wavelength of 1.55~$\mu$m, the quantum efficiency of the camera is constant with the avalanche gain as shown in Figure 6 of \cite{gert2016}. We double checked the temporal stability of the flux using a calibrated diode. Over the course of the experiment, the calibrated photometry of our light source was stable within a few percents. 

\begin{figure}[h]
\centering
\includegraphics[scale=0.54]{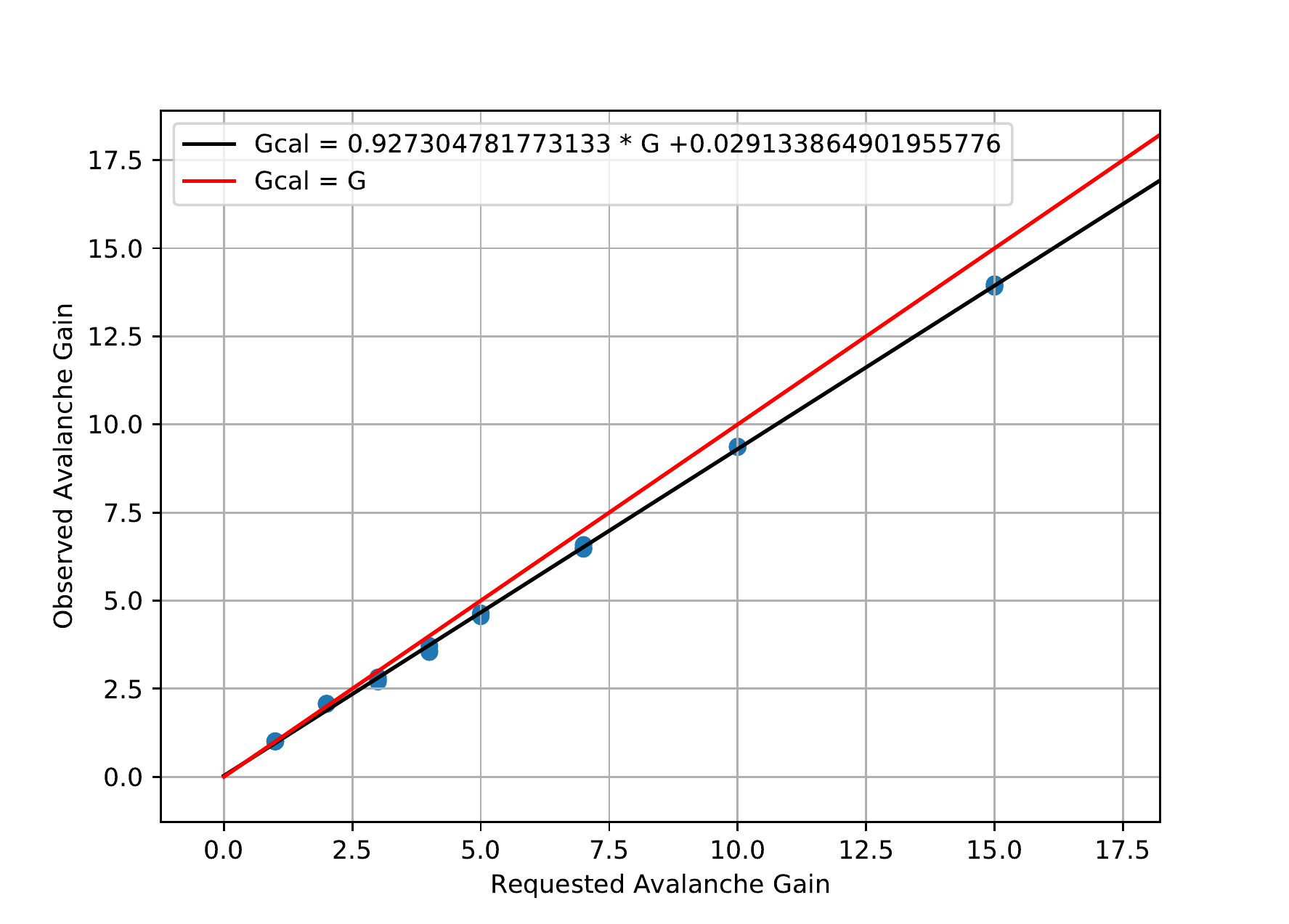}
\caption{Derived avalanche gain as a function of the avalanche gain provided by the manufacturer. The red line is the 1:1 relation. The black line is a linear fit, excluding the unity gain.}\label{fig:GcalGav}
\end{figure}

Figure~\ref{fig:GcalGav} displays the measured avalanche gain compared to the manufacturer calibration. The unity gain is used as a reference to compute the avalanche gain and therefore lies on the red curve by definition. Our measured avalanche gain is consistently $0.93$ times the avalanche gain calibrated by the manufacturer. Although not presented in Fig.~\ref{fig:GcalGav}, we have verified that this relation applies for avalanche gain up to 150.

The calibration from the manufacturer is based on a different method that is a mean-variance analysis for each amplification gain. Our independent measurement confirms this calibration. The remaining discrepancy of a few percents on the absolute gain is not an issue in our operational scenario. Nor the real-time fringe tracking algorithms neither the instrument data reduction algorithms rely on an absolute knowledge of the gain.

\subsection{Attempt of photon counting}\label{sec:phcouttry}

\begin{figure}[!ht]
\centering
\includegraphics[scale=0.57]{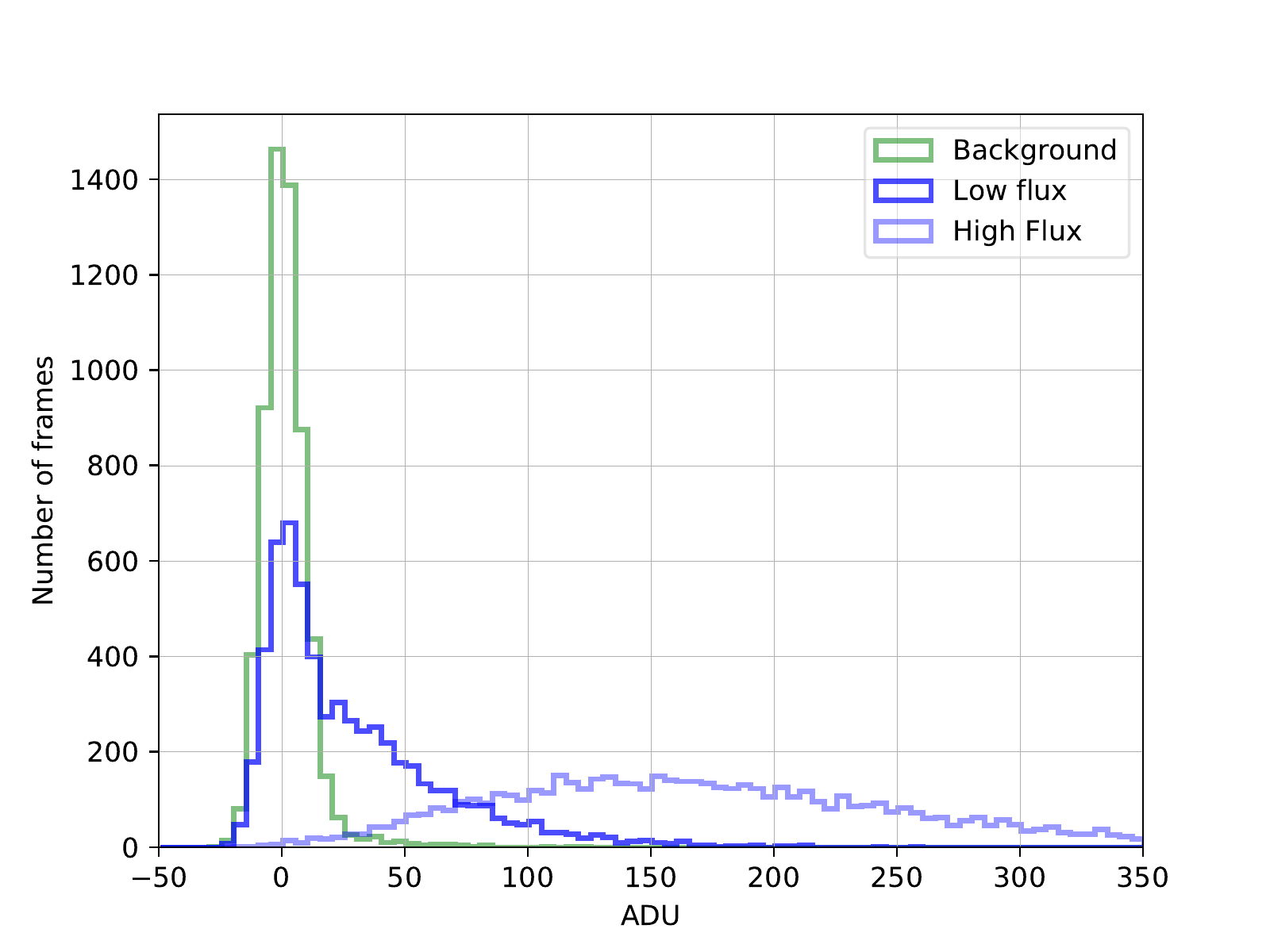}
\caption{Histograms of counts measured on a single pixel, with a low-flux illumination ($<1$~e/frame/pixel, darker blue) and with a high flux illumination ($>3$~e/frame/pixel, lighter blue). The total gain is 50~$\ADU/$e. The histogram without illumination is shown in green and represents the intrinsic statistic of the 0-photon events. \label{fig:histogram}}
\end{figure}

For a total gain of $50\,\ADU/$e or higher, the signal of incoming individual photons should overwhelm the readout noise of $12\,\ADU$ and therefore allows photon counting. This is especially true since the expected excess noise factor (\ENF{}) is small (1.25). The ENF is the noise added by the avalanche process of the e-APD. The lower it is, the lower the avalanche process add noise. It has a limit of 1 for noiseless avalanche process. The effect of the ENF on the photon counting is that it spread the signal distribution of one photon. To explore this possibility of individual photon counting, we compute the histograms of the temporal sequences of the values of a single illuminated pixel. Typical histograms are shown in Fig.~\ref{fig:histogram}. We interpret the histogram of the background as the sum of (1) a Gaussian distribution of RMS 12~\ADU{} corresponding to the detector readout noise; and (2) a small tail toward positive values corresponding to the few frames with a background photon (approximately one over 23 frames). The histogram measured in the low flux regime shows a break at 25 \ADU{}. We interpret this break as the transition between the 0-photon events and the 1 or more photon events. It means that we can derive the proportion of 0-photon events, hence the incoming flux. In the high flux regime, the fraction of 0-photon event becomes negligible. The clear transition between the 0-photon event and the 1- or more photons events is not noticeable anymore. Therefore, it appears that it is not possible to perform discretized photon-counting with this camera. In the following, we develop a simple model that reproduces these histograms and allows us to measure the excess noise factor.


\section{Modeling the photonic signal}\label{sec:dataproc}

We record histograms for various illuminations and various gains. Because of the very low flux levels we are interested in, it was unfortunately impossible to build an absolute calibration of the illumination. 

\subsection{Model}
\label{sec:model}
In the following, we call \Flux{} the mean flux reaching the considered pixel and coming from the internal source, in e/frame/pixel. This is an unknown parameter. $\Flux{}=0$ is, by definition, the mean of the histogram of the background counts (in green in Fig.~\ref{fig:histogram}).
We developed a simple model with the goal to reproduce the observed distributions $\mathrm{H}(\ADU)$ as a function of the parameters \Gain{} (gain in ADU/photon), the \Flux{} (flux in photon/frame/pixel) and the \ENF{} (excess noise factor). Our model is described by the following equation:
\begin{equation}\label{eq:model}
\mathrm{H} = \overbrace{BKG(\ADU,\Gain)}^{(3)}\;\;*\;\;\overbrace{\sum_{p} \bigg[P(p, \Flux)}^{(2)}\;\cdot\;\overbrace{MC(\ENF, \frac{\ADU}{\Gain})}^{(1)}\bigg],
\end{equation}
where the $*$ sign is for convolution, the $\cdot$ sign is for multiplication, $BKG(\ADU,\Gain)$ is the measured distribution without illumination, $P(p, \Flux)$ is the Poisson probability of a $p$-photon event given a mean flux  Flux{}, and 
\begin{equation}
MC = \overbrace{M(\ENF, \ADU/\Gain) * M(\ENF, \ADU/\Gain) * ...}^{p - 1\; \mathrm{convolutions}}
\end{equation}
is the iterative self-convolution of the gain distribution because each avalanche generates its own avalanche gain distribution. This model is based on the following realistic assumptions and follows these steps:
\begin{enumerate}
    \item The amplification gains of the individual photons summed up in a frame with $p$-photon are not correlated. Therefore the distribution of the $p$-photon frames is represented by $p - 1$ convolutions of the gain distribution $M$.
    \item These $p$-photon frames are uncorrelated, therefore the combined distribution is a weighted sum, according to the Poisson distribution for the desired flux.
    \item The background flux is additive and uncorrelated with the illumination flux. Therefore the distribution given by the previous steps is convoluted by the distribution measured on the background.
\end{enumerate}
The key physical ingredient of this model is the gain distribution $M$.

\subsection{Prescription for the gain distribution}

The mean gain \Gain{} and the excess noise factor \ENF{} are intrinsic properties of the gain distribution $M$. A Gaussian distribution~\citep{n-photon} truncated over $\mathbb{R}^+$ can only model low \ENF{} values ($<$ 1.2). This is because the fraction of non-physical ``negative amplifications'' becomes numerically significant at larger ENF values. A decreasing exponential distribution truncated over $\mathbb{R}^+$~\citep{4711091} has a fixed \ENF{} of 2.0. It is unsuited to reproduce the expected \ENF{} of 1.25 of our camera.

The Gamma distribution is classically used to model the photon distribution in EMCCD cameras~\citep{EMCCDgamma}. It describes the process of multiplication of a single electron inside the pixel. The Gamma distribution is an intrinsically asymmetric function defined and normalized over $\mathbb{R}^+$. It is defined by 2 positive parameters $k$ and $\theta$  (see Appendix~\ref{an:gammadist}). The former parameter describes the shape of the function, while the latter describes its scale. These 2 parameters can be directly linked to the ENF and the mean gain by:
\begin{equation}\label{eq:k}
k = \frac{1}{\ENF-1}
\end{equation}
and
\begin{equation}\label{eq:theta}
\theta = G ( \ENF - 1).
\end{equation}

The Gamma distribution allows spanning the entire range of ENF values between 1 and $+\infty$. Figure~\ref{fig:Gdist} displays four examples of the Gamma distribution. Interestingly, the exponential distribution and the Gaussian distribution are special cases of the Gamma distribution. Therefore these two  situations are included in our parameter space.

\subsection{Fit strategy}\label{sec:Fit}

Our model uses three free parameters (the mean gain \Gain{}, the excess noise factor \ENF{}, and the mean photonic flux $F$) in order to reproduce the observed distribution $\mathrm{H}(\ADU)$. The background distribution $BKG(\ADU)$ is measured by turning off the internal source.

We first run a brute-force minimization by computing the reduced chi-squared \chis{} between the data and the model for each point in the cube (\Gain{}, \ENF{}, $F$). At the end of the brute force method, we run a classical gradient descent toward the best-fit parameters with the minimum \chis{} position in the cube as the starting point. The whole process on the 25 data files takes $\approx$3 hours with 25 points on the grid for each parameter. We checked that this sampling is enough to sample the degeneracy by processing some data sets with 50 and 100 points on the grid for each parameter.


\section{Results and discussion}\label{sec:results}

\subsection{Best-fit quality}\label{sec:best-fit}

\begin{figure*}[!ht]
\centering
\includegraphics[scale=0.9]{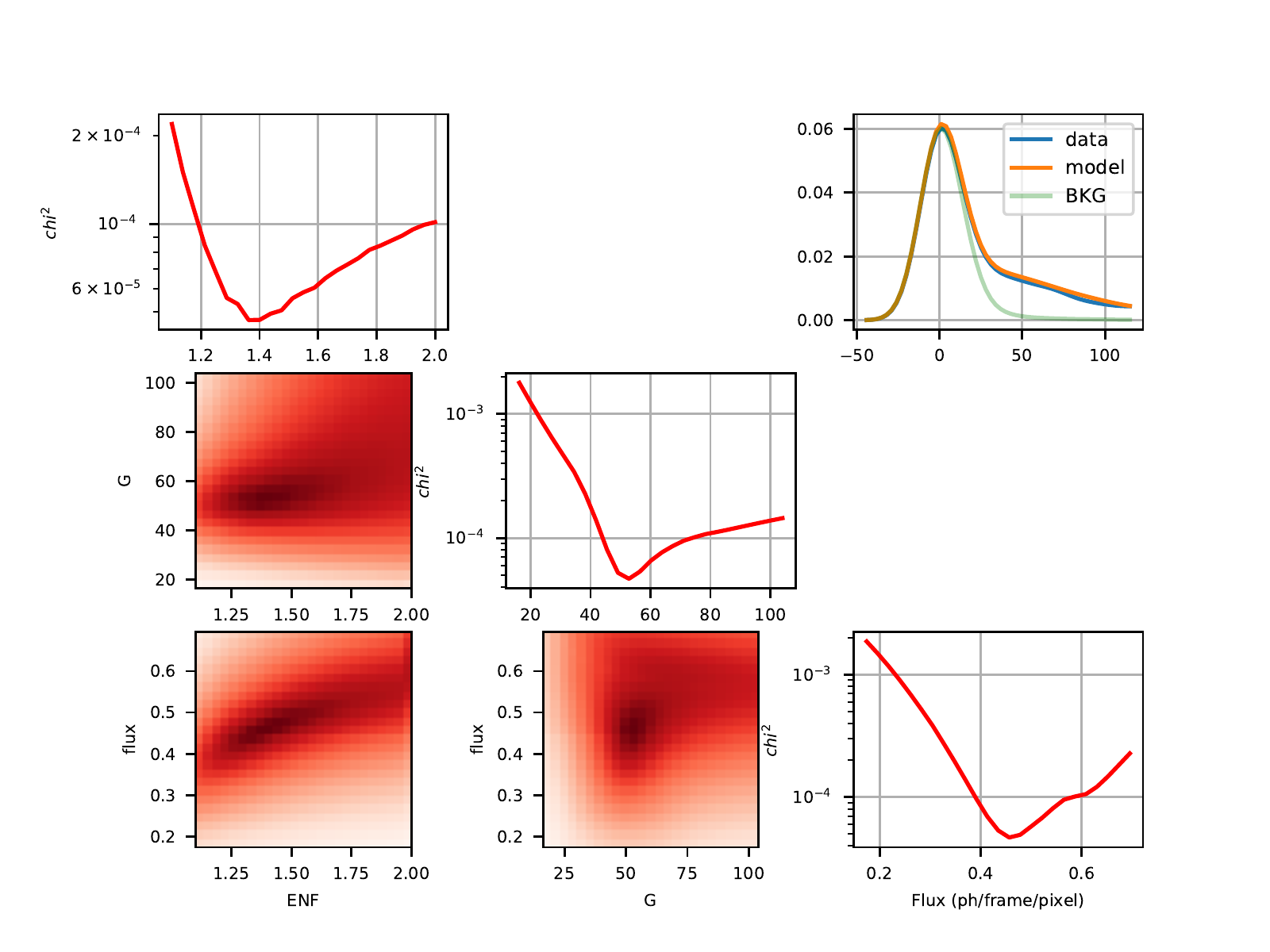}
\caption{Example of \chis{} minimization result. This example is from data with Gman = 76.7, the expected gain from the manufacturer calibration. The parameters that minimize the \chis{} are \Flux{} = 0.46, \Gain{} = 52.2 and \ENF{} = 1.38. {\bf Diagonal.} Minimum \chis{} as a function of \ENF{} (top left), \Gain{} (middle) and \Flux{} (bottom right), the two other parameters being free. {\bf Lower left.} Maps of minimum \chis{} as a function of two parameters. {\bf Upper right.} Histogram for the data with illumination in blue, histogram of the background in light green, and histogram obtained with the parameters that minimize the \chis{} in orange.}\label{fig:chi2map1}
\end{figure*}

Fig.~\ref{fig:chi2map1} shows the results of the minimization for a data set with a low flux level. The overall shape of the data is convincingly reproduced by the best-fit solution. The solution has no degeneracy, there is only one minimum for each parameter. This is because the fraction of 0-photon events is well constrained by the "bump" at 0. Knowing the flux, the mean on the histogram constrains the gain. Then, the ENF is constrained by the shape of the histogram tail.

Appendix~\ref{an:minimres} shows all the results. The best-fit models reproduce convincingly the data for all flux and gain. This confirms that the Gamma distribution is an adequate model of the amplification process. As expected, for $\Flux>3$\, e/frame/pixel, the degeneracy of the model increased significantly, as shown on Fig~\ref{fig:chi2map2-lghf}, \ref{fig:chi2map2-hghf} and \ref{fig:chi2map2-200}. This is due to the fact that the proportion of 0 photon events is not significant enough anymore to constrain the flux with the data histograms. The \chis{} maps show that we have only a lower limit to \Flux{}. As a consequence, we only have a higher limit for \Gain{}, and \ENF{} is poorly constrained. 

\subsection{Gain}\label{sec:Gain}

\begin{figure}[!ht]
\centering
\includegraphics[scale = 0.6]{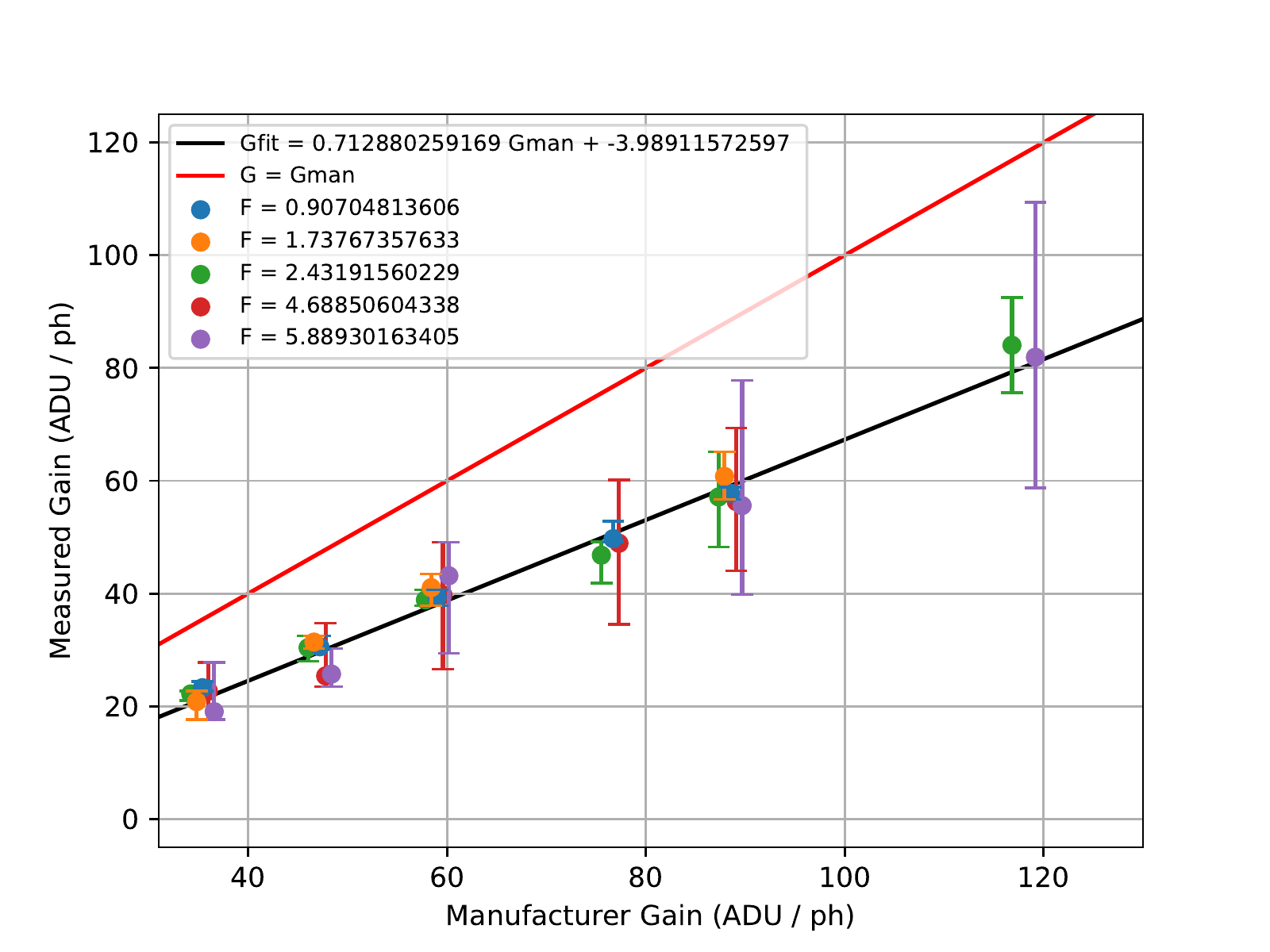}
\caption{Best-fit gain as a function of the expected gain from the manufacturer calibration. The black line is a linear fit, while the red line is the expected 1:1 relation. The different colors of the data points show the different flux estimated by the model-fitting.}\label{fig:GGerrbar}
\end{figure}

Figure~\ref{fig:GGerrbar} displays the best-fit gain versus the expected gain as calibrated by the manufacturer. The results obtained with different flux levels and gains are self-consistent: our best-fit gain (\Gain{}) is consistent $0.71$ times the expected gain from the manufacturer calibration ($\Gain_{man}$). We notice that the higher the flux level, the larger the uncertainties. This is due to the degeneracy at high flux levels. Interestingly, the best-fit gains for those high flux data are still consistent with the ones at low flux.

We analyzed several pixels in order to check the uniformity of the results in the matrix (see Appendix~\ref{an:diffpix}). The results are consistent over the pixels and show a mean gain $\Gain = 0.70 \pm 0.04 \times \Gain_{man}$. This is in overall agreement with the fully independent calibrations presented in Sec.~\ref{sec:detectcharac}. With these more classical methods, we obtained a ratio of $G_{\rm sys\, measured} / G_{\rm sys\, man} \times G_{\rm ava\, ratio} = 0.49 / 0.59 \times 0.93 = 0.77$ for the total gain, compared to the manufactured calibration. This also confirms that our simple model is adequately capturing the statistic of the amplification process.

\subsection{Excess Noise Factor}\label{sec:ENF}

\begin{figure}[!ht]
\centering
\includegraphics[scale = 0.6]{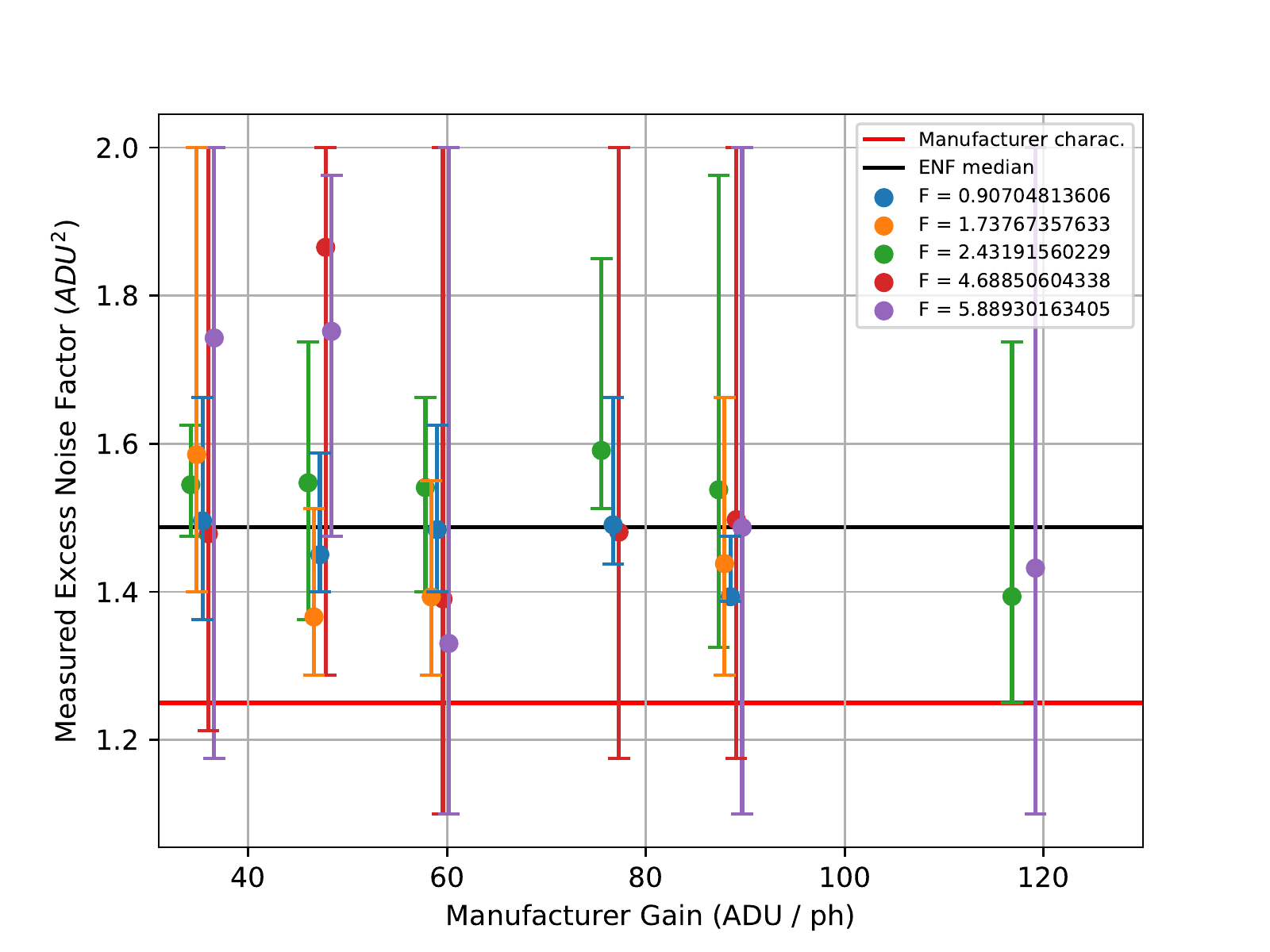}
\caption{Best-fit ENF as a function of the avalanche gain. Colors represent the flux estimated by the best-fit. The black line is the median value of measurements. The red line is the value from the manufacturer.}\label{fig:ENFerrbar}
\end{figure}

Figure~\ref{fig:ENFerrbar} shows the measured \ENF{} as a function of the avalanche gain. The median value of all the measured \ENF{} is $1.49 \pm 0.12$. As expected, the uncertainty increases for measurements at higher flux but the error bars all remain compatible with the median value. As for the gain, we performed the same study on several pixels. The results are summarized in Appendix~\ref{an:diffpix}. The mean \ENF{} over the pixels is $1.47 \pm 0.03$.

The discrepancy between our measured \ENF{} and the value reported in the literature is striking. \cite{gert2016} measure an \ENF{} of 1.3 at a temperature of 90~K and 1.1 at 40~K. Their method to measure the \ENF{}  is independent of a gain distribution model. They control the illumination by using a black body source located in front of the detector. They compare the noise on the detector with the photon noise expected for the controlled flux. Based on this calibration, the value expected by the manufacturer for our camera is 1.25 for an operating temperature of 80~K.

The detailed study of a MARK13 SAPHIRA presented by \citet{2018AJ....155..220A} gives interesting clues on the \ENF{}. Their Figure~6 shows a histogram of the detector signal with very low illumination (about 0.1~photon/frame/pixel). The histogram has a peak at the position of the 0-photon events whose shape is similar to the histogram without illumination,  and a tail for the one or more photons events. As in our measurements, there is no separation between those two distributions. We run our model on the data extracted from Figure~6 of \citet{2018AJ....155..220A}. The results are presented in Fig.~\ref{fig:chi2map2-atk}. The model reproduces adequately the shape of the observed histogram. The best fit \ENF{} = 1.45 is consistent with the results obtained on our own data.

The first possibility is the somewhat arbitrary choice of the gain distribution model $M$ in our study. We also tried a Gaussian model defined over $\mathbb{R}^+$. The resulting \ENF{} were not significantly different. 

Another possibility is that the illumination contains photons at higher wavelengths. These photons are absorbed deeper in the pixel, possibly inside the multiplication region. This would create a less amplified signal, spreading the gain distribution. But our optical setup disperses the spectrum of the white-light source. Based on the spectral calibration of the instrument, we are sure that the pixels considered for this study only receive photons between 1.55 and 1.6 $\mu$m coming from the white-light source. Remember that the photons of the warm background are already taken into account in the background histogram.

Another hypothesis is that the signal of individual photon is temporally smeared. In \cite{gert2016}, the detector response time is of the order of 1\,ms. As the measurement of the ENF in the same paper is made at 1\,kHz, the signal would not be smeared. In our experiment, we took data at 1900\,Hz, which is faster than the detector response time. But the claim is that the detector response time is dominated by the electron migration in the substrate and not by the avalanche process \citep{gert2016}. Hence, it would delay the signal, not smear it. Another smearing occurs due to our specific readout mode, for photons arriving between consecutive readout that are averaged together into a single frame. We read an entire row several times before reading the others, and average these measurements. As we read 20 rows, it means that only 5$\%$ of the photons are smeared this way. This is too low to explain the difference between our value and the ones in the literature. Moreover, such an effect skews the distribution of gain toward the lower values (negative skewness), not toward the higher values (positive skewness) as seen in Figure~6 of \citet{2018AJ....155..220A} and across this paper.

The last possibility is that the results existing in the literature have been obtained for relatively high flux (several tens of photons per frame and per pixel), while our study and the one of Atkinson focuses on very low flux levels (less than 5 photons per frame and per pixel, and more typically less than 1 photon per frame and per pixel). We suspect that the physics of the detector is not the same when there is one or several electrons migrating in the amplification zone.

\subsection{Consequences for photon-counting}\label{sec:threshold}

As shown in Sect.~\ref{sec:phcouttry}, photon counting by fully separating the distributions of zero, single, or multiple photon events is not possible with our camera. This is due to the unexpected \ENF{} of 1.47. However, Fig.~\ref{fig:histogram} shows that we do see two different distributions at low flux: the one of the 0-photon events, and the one of the 1 or more photon events. Therefore, one could define a threshold that would separate, with some statistical uncertainties, the frames with and without photons. Computing this optimal threshold for each gain and each frame rate is out of the scope of this paper.

Although disentangling the 0- and the $\geq$$1$-photon events may appear attractive, we did not implement it in the data reduction pipeline of the instrument. Indeed, when dealing with a significant fraction of $\geq$$2$-photon events, such threshold dramatically complicates the un-biasing of the fringe bi-spectrum. Detailed discussion on the impact of the measurement statistic in the fringe bi-spectrum can be found in \citet{2004MNRAS.347.1187B}, \citet{2012A&A...541A..46G} and \citet{2016PASP..128e5004G}. We are currently upgrading the bi-spectrum de-biasing algorithm in the MIRCX pipeline based on our new understanding of the signal statistic presented in this study.

\subsection{Consequences on instrument performance}
\label{sec:comparison}

\begin{figure}[!ht]
\centering
\includegraphics[width=\columnwidth]{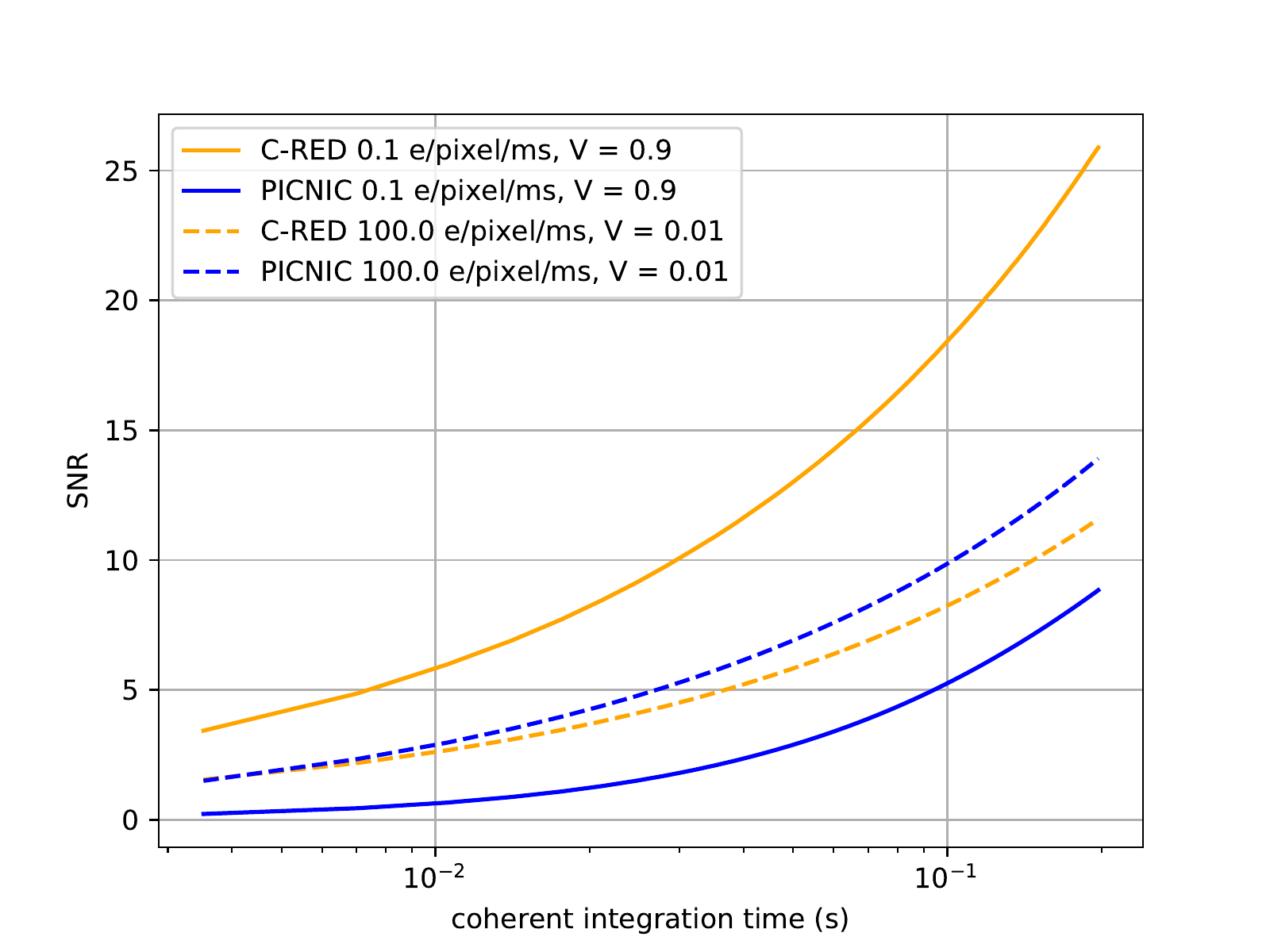}
\caption{Simulation of the SNR as a function of the coherent integration time in two operational conditions of flux and fringe visibility, for the previous PICNIC camera of MIRC and the new C-RED ONE camera of MIRC-X.}\label{fig:SNRPICMIRCX}
\end{figure}

We simulate the SNR of the MIRC-X instrument for two realistic situations. First, we simulate a flux of 0.1~e/pixel/ms which is typically obtained on faint objects, such as Young Stellar Objects. These are the primary science goals of MIRC-X. The fringe visibility is often high (close to 1) because the objects are unresolved. Secondly, we simulate a flux of 100~e/pixel/ms which is typical of bright objects, such as fully resolved Stellar Surfaces. These targets have very low visibility (e.g. 0.01). We estimate the detector and background noise by using actual measurements in background exposures obtained with the former PICNIC camera and with the C-RED ONE at an avalanche gain of 60 \citep[best compromise between readout noise and tunneling noise,][]{SPIE2018}. We used the measured \ENF{} = 1.47 for the C-RED ONE camera, and ENF = 1 for the PICNIC camera, as this one doesn't have additional noise from an avalanche process. For both cameras, the fringe signal is encoded into 100 pixels. Results are shown in Fig.~\ref{fig:SNRPICMIRCX}.

The operational limit for fringe detection is defined by $\mathrm{SNR}>3$. In the low flux regime, the simulation shows an improvement of one decade in term of necessary integration time between the former PICNIC camera and the new C-RED ONE. It corresponds to a gain in sensitivity of 2.5 magnitudes. In the high flux regime, the PICNIC camera is more efficient but the difference is rather small in term of limiting magnitude. For the specific case of averaging many exposures on bright targets with the goal to reach a  high SNR, the PICNIC camera remains theoretically  advantageous because it doesn't suffer from the Excess Noise Factor. However, this is not true in practice because other effects quickly become dominant such as the piston noise or the calibration accuracy.

This simulation compares realistically with the results obtained on sky so far. The typical limiting magnitude of the instrument with the former PICNIC camera was $\mathrm{H} \approx 5.5$. Since the upgrade, it has been possible to observe several Young Stellar Objects and massive stars at magnitudes $\mathrm{H}\approx7.5$. Overall, in term of scientific productivity for MIRC-X, the sensitivity boost obtained in the low flux regime largely overcome the possible sensitivity losses in the high flux regime.

\section{Conclusions}\label{sec:concl}

We have characterized the first near-infrared C-RED ONE camera sold by the First Light Imaging company. This SAPHIRA electron avalanche photo-diode camera is now part of the \mbox{MIRC-X} instrument operating in the H band and installed at the combined focus of the CHARA array.

We have presented the characterization of this C-RED ONE camera in the operational scenario of the MIRC-X instrument (multiple readouts, kHz frame rate and low flux illumination). We first performed a classical analysis based on the Mean-Variance curve and mean flux level in order to determine the system gain and amplification gain. We have then presented a simple, but innovative, modeling of the signal distribution from the camera at low flux level. It enabled us to estimate both the total gain and the excess noise factor brought by the amplification process. The model reproduces adequately the observed distributions of photon events for a large range of flux and gains. We have also applied our model to existing literature data to confirm our measurements.

We measured a system gain and total gain slightly lower than the ones calibrated by the manufacturer. This result is observed consistently when using the Mean-Variance method and our model based on the signal distribution. Our modeling also revealed that the \ENF{} (1.47) is significantly higher than the values in the literature for this type of detectors (1.25). We could not determine the reason for such a discrepancy. After reviewing various possibilities, we proposed that the amplification process could behave differently in the low flux regime and high flux regime. 

From section~\ref{sec:comparison},  we show that the C-RED ONE in the MIRC-X instrument brings a theoretical improvement in sensibility of 2.5 magnitudes in the H band for faint objects, when its performances are comparable to the former \mbox{PICNIC} camera for bright objects.
More generally, with $<0.3$ electron equivalent readout noise at kilohertz frame rates computed from the background histograms (Fig~\ref{fig:histogram}), we confirm the revolutionary performances of the camera with respect to the PICNIC or HAWAII technologies for high frame rate applications. In current operation, the camera has improved the limiting magnitude of our MIRC-X instrument by up to two magnitudes, the exact value depending on the atmospheric coherence time and the setup of the spectrograph. However, the observed higher excess noise factor is consistent with the fact that we cannot separate the events with different numbers of photons in the observed signal distribution.

\begin{acknowledgements} 

We would like to thanks Marc-Antoine MARTINOD, Johan ROTHMAN and Romain LAUGIER for their interest in this work and the fruitful discussions that lead to the choice of the Gamma distribution for the gain distribution. We thank also FLI for their collaboration in the characterization and the comprehension of the camera. We also thank the University of Michigan to welcome us and to make possible this collaboration work. We finally thank CHARA and all the people that allows us to take data and observe. This work has been partially supported by the LabEx FOCUS ANR-11-LABX-0013. The research leading to these results has received funding from the European Unions Horizon 2020 research and innovation programme under Grant Agreement 730890 (OPTICON) and Grant Agreement No. 639889 (ERC Starting Grant ”ImagePlanetFormDiscs”). This work has been supported by a grant from Labex OSUG@2020 (Investissements d’avenir ANR10 LABX56). This work was supported by the Programme National de Physique Stellaire (PNPS) of CNRS/INSU co-funded by CEA and CNES. This work was supported by the Action Spécifique Haute Résolution Angulaire (ASHRA) of CNRS/INSU co-funded by CNES.
\end{acknowledgements}

\bibliographystyle{aa}
\bibliography{biblio}

\begin{thebibliography}{29}
\expandafter\ifx\csname natexlab\endcsname\relax\def\natexlab#1{#1}\fi

\bibitem[{{Atkinson} {et~al.}(2018){Atkinson}, {Hall}, {Jacobson}, \&
  {Baker}}]{2018AJ....155..220A}
{Atkinson}, D., {Hall}, D., {Jacobson}, S., \& {Baker}, I.~M. 2018, \aj, 155,
  220

\bibitem[{{Basden} \& {Haniff}(2004)}]{2004MNRAS.347.1187B}
{Basden}, A.~G. \& {Haniff}, C.~A. 2004, \mnras, 347, 1187

\bibitem[{{Beckmann} {et~al.}(2014){Beckmann}, {Connot}, {Heininger},
  {Hofmann}, {Nu{\ss}baum}, {Schertl}, {Solscheid}, {ten Brummelaar}, {Turner},
  \& {Weigelt}}]{2014SPIE.9146E..1WB}
{Beckmann}, U., {Connot}, C., {Heininger}, M., {et~al.} 2014, in \procspie,
  Vol. 9146, Optical and Infrared Interferometry IV, 91461W

\bibitem[{Bohndiek {et~al.}(2008)Bohndiek, Blue, Clark, Prydderch, Turchetta,
  Royle, \& Speller}]{PTC}
Bohndiek, S.~E., Blue, A., Clark, A.~T., {et~al.} 2008, IEEE Sensors Journal,
  8, 1734

\bibitem[{{Feautrier} {et~al.}(2014){Feautrier}, {Gach}, {Guieu}, {Downing},
  {Jorden}, {Rothman}, {de Borniol}, {Balard}, {Stadler}, {Guillaume},
  {Boutolleau}, {Coussement}, {Kolb}, {Hubin}, {Derelle}, {Robert}, {Tanchon},
  {Trollier}, {Ravex}, {Zins}, {Kern}, {Moulin}, {Rochat}, {Delpoulb{\'e}}, \&
  {Lebouqun}}]{2014SPIE.9148E..18F}
{Feautrier}, P., {Gach}, J.-L., {Guieu}, S., {et~al.} 2014, in \procspie, Vol.
  9148, Adaptive Optics Systems IV, 914818

\bibitem[{Finger {et~al.}(2016)Finger, Baker, Alvarez, Dupuy, Ives, Meyer,
  Mehrgan, Stegmeier, \& Weller}]{gert2016}
Finger, G., Baker, I., Alvarez, D., {et~al.} 2016, Proc.SPIE, 9909, 9909

\bibitem[{Finger {et~al.}(2012)Finger, Baker, Alvarez, Ives, Mehrgan, Meyer,
  Stegmeier, Thorne, \& Weller}]{gert2012}
Finger, G., Baker, I., Alvarez, D., {et~al.} 2012, Proc.SPIE, 8453, 8453

\bibitem[{Finger {et~al.}(2010)Finger, Baker, Dorn, Eschbaumer, Ives, Mehrgan,
  Meyer, \& Stegmeier}]{gert2010}
Finger, G., Baker, I., Dorn, R., {et~al.} 2010, Proc.SPIE, 7742, 7742

\bibitem[{{Garcia} {et~al.}(2016){Garcia}, {Muterspaugh}, {van Belle},
  {Monnier}, {Stassun}, {Ghasempour}, {Clark}, {Zavala}, {Benson}, {Hutter},
  {Schmitt}, {Baines}, {Jorgensen}, {Strosahl}, {Sanborn}, {Zawicki},
  {Sakosky}, \& {Swihart}}]{2016PASP..128e5004G}
{Garcia}, E.~V., {Muterspaugh}, M.~W., {van Belle}, G., {et~al.} 2016, \pasp,
  128, 055004

\bibitem[{{Gies} {et~al.}(2018){Gies}, {ten Brummelaar}, {Anderson},
  {Farrington}, {Golden}, {Jones}, {Klement}, {Majoinen}, {Schaefer},
  {Sturmann}, {Sturmann}, {Turner}, {Vargas}, {Webster}, {Woods}, \&
  {Ridgway}}]{2018SPIE10701E..02G}
{Gies}, D.~R., {ten Brummelaar}, T.~A., {Anderson}, M.~D., {et~al.} 2018, in
  Society of Photo-Optical Instrumentation Engineers (SPIE) Conference Series,
  Vol. 10701, Society of Photo-Optical Instrumentation Engineers (SPIE)
  Conference Series, 1070102

\bibitem[{{Gordon} \& {Buscher}(2012)}]{2012A&A...541A..46G}
{Gordon}, J.~A. \& {Buscher}, D.~F. 2012, \aap, 541, A46

\bibitem[{{GRAVITY collaboration} {et~al.}(2017){GRAVITY collaboration},
  {Abuter}, {Accardo}, {Amorim}, {Anugu}, {{\'A}vila}, {Azouaoui}, {Benisty},
  {Berger}, {Blind}, {Bonnet}, {Bourget}, {Brandner}, {Brast}, {Buron},
  {Burtscher}, {Cassaing}, {Chapron}, {Choquet}, {Cl{\'e}net}, {Collin},
  {Coud{\'e} Du Foresto}, {de Wit}, {de Zeeuw}, {Deen},
  {Delplancke-Str{\"o}bele}, {Dembet}, {Derie}, {Dexter}, {Duvert}, {Ebert},
  {Eckart}, {Eisenhauer}, {Esselborn}, {F{\'e}dou}, {Finger}, {Garcia}, {Garcia
  Dabo}, {Garcia Lopez}, {Gendron}, {Genzel}, {Gillessen}, {Gonte}, {Gordo},
  {Grould}, {Gr{\"o}zinger}, {Guieu}, {Haguenauer}, {Hans}, {Haubois}, {Haug},
  {Haussmann}, {Henning}, {Hippler}, {Horrobin}, {Huber}, {Hubert}, {Hubin},
  {Hummel}, {Jakob}, {Janssen}, {Jochum}, {Jocou}, {Kaufer}, {Kellner},
  {Kendrew}, {Kern}, {Kervella}, {Kiekebusch}, {Klein}, {Kok}, {Kolb}, {Kulas},
  {Lacour}, {Lapeyr{\`e}re}, {Lazareff}, {Le Bouquin}, {L{\`e}na}, {Lenzen},
  {L{\'e}v{\^e}que}, {Lippa}, {Magnard}, {Mehrgan}, {Mellein}, {M{\'e}rand},
  {Moreno-Ventas}, {Moulin}, {M{\"u}ller}, {M{\"u}ller}, {Neumann}, {Oberti},
  {Ott}, {Pallanca}, {Panduro}, {Pasquini}, {Paumard}, {Percheron}, {Perraut},
  {Perrin}, {Pfl{\"u}ger}, {Pfuhl}, {Phan Duc}, {Plewa}, {Popovic}, {Rabien},
  {Ram{\'\i}rez}, {Ramos}, {Rau}, {Riquelme}, {Rohloff}, {Rousset}, {Sanchez-
  Bermudez}, {Scheithauer}, {Sch{\"o}ller}, {Schuhler}, {Spyromilio},
  {Straubmeier}, {Sturm}, {Suarez}, {Tristram}, {Ventura}, {Vincent},
  {Waisberg}, {Wank}, {Weber}, {Wieprecht}, {Wiest}, {Wiezorrek}, {Wittkowski},
  {Woillez}, {Wolff}, {Yazici}, {Ziegler}, \& {Zins}}]{2017A&A...602A..94G}
{GRAVITY collaboration}, {Abuter}, R., {Accardo}, M., {et~al.} 2017, \aap, 602,
  A94

\bibitem[{{Greffe} {et~al.}(2016){Greffe}, {Feautrier}, {Gach}, {Stadler},
  {Clop}, {Lemarchand}, {Boutolleau}, \& {Baker}}]{2016SPIE.9907E..2EG}
{Greffe}, T., {Feautrier}, P., {Gach}, J.-L., {et~al.} 2016, in Optical and
  Infrared Interferometry and Imaging V, Vol. 9907, 99072E

\bibitem[{{Guieu} {et~al.}(2014){Guieu}, {Feautrier}, {Zins}, {Le Bouquin},
  {Stadler}, {Kern}, {Rothman}, {Tauvy}, {Coussement}, {de Borniol}, {Gach},
  {Jacquard}, {Moulin}, {Rochat}, {Delboulb}, {Derelle}, {Robert},
  {Vuillermet}, {M{\'e}rand}, \& {Bourget}}]{2014SPIE.9146E..1NG}
{Guieu}, S., {Feautrier}, P., {Zins}, G., {et~al.} 2014, in Optical and
  Infrared Interferometry IV, Vol. 9146, 91461N

\bibitem[{{Haguenauer} {et~al.}(2012){Haguenauer}, {Abuter}, {Andolfato},
  {Alonso}, {Blanchard}, {Berger}, {Bourget}, {Brillant}, {Derie},
  {Delplancke}, {Di Lieto}, {Dupuy}, {Gilli}, {Gitton}, {Gonzalez}, {Guisard},
  {Guniat}, {Hudepohl}, {Kaufer}, {L{\'e}v{\^e}que}, {M{\'e}nardi},
  {M{\'e}rand}, {Morel}, {Percheron}, {Phan Duc}, {Poupar}, {Ramirez},
  {Reineiro}, {Rengaswamy}, {Rivinius}, {Sch{\"o}ller}, {Schmid}, {Segovia},
  {Schuhler}, {Valdes}, {de Wit}, \& {Wittkowski}}]{2012SPIE.8445E..0DH}
{Haguenauer}, P., {Abuter}, R., {Andolfato}, L., {et~al.} 2012, in \procspie,
  Vol. 8445, Optical and Infrared Interferometry III, 84450D

\bibitem[{Hirsch {et~al.}(2013)Hirsch, Wareham, Martin-Fernandez, Hobson, \&
  Rolfe}]{EMCCDgamma}
Hirsch, M., Wareham, R.~J., Martin-Fernandez, M.~L., Hobson, M.~P., \& Rolfe,
  D.~J. 2013, PLOS ONE, 8, 1

\bibitem[{Kardyna{\l} {et~al.}(2008)Kardyna{\l}, Yuan, \& Shields}]{n-photon}
Kardyna{\l}, B.~E., Yuan, Z.~L., \& Shields, A.~J. 2008, Nature Photonics, 2,
  425 EP

\bibitem[{Kraus {et~al.}(2018)Kraus, Monnier, Anugu, Bouquin, Davies, Ennis,
  Labdon, Lanthermann, Setterholm, \& ten Brummelaar}]{MIRCX}
Kraus, S., Monnier, J.~D., Anugu, N., {et~al.} 2018, Proc.SPIE, 10701, 10701

\bibitem[{Lanthermann {et~al.}(2018)Lanthermann, Bouquin, Anugu, Monnier, \&
  Kraus}]{SPIE2018}
Lanthermann, C., Bouquin, J.-B.~L., Anugu, N., Monnier, J., \& Kraus, S. 2018,
  Proc.SPIE, 10709, 10709

\bibitem[{{Le Bouquin} {et~al.}(2011){Le Bouquin}, {Berger}, {Lazareff},
  {Zins}, {Haguenauer}, {Jocou}, {Kern}, {Millan-Gabet}, {Traub}, {Absil},
  {Augereau}, {Benisty}, {Blind}, {Bonfils}, {Bourget}, {Delboulbe},
  {Feautrier}, {Germain}, {Gitton}, {Gillier}, {Kiekebusch}, {Kluska},
  {Knudstrup}, {Labeye}, {Lizon}, {Monin}, {Magnard}, {Malbet}, {Maurel},
  {M{\'e}nard}, {Micallef}, {Michaud}, {Montagnier}, {Morel}, {Moulin},
  {Perraut}, {Popovic}, {Rabou}, {Rochat}, {Rojas}, {Roussel}, {Roux},
  {Stadler}, {Stefl}, {Tatulli}, \& {Ventura}}]{2011A&A...535A..67L}
{Le Bouquin}, J.-B., {Berger}, J.-P., {Lazareff}, B., {et~al.} 2011, \aap, 535,
  A67

\bibitem[{Monnier {et~al.}(2018)Monnier, Bouquin, Anugu, Kraus, Setterholm,
  Ennis, Lanthermann, Jocou, \& ten Brummelaar}]{MYSTIC}
Monnier, J.~D., Bouquin, J.-B.~L., Anugu, N., {et~al.} 2018, Proc.SPIE, 10701,
  10701

\bibitem[{{Monnier} {et~al.}(2006){Monnier}, {Pedretti}, {Thureau}, {Berger},
  {Millan-Gabet}, {ten Brummelaar}, {McAlister}, {Sturmann}, {Sturmann},
  {Muirhead}, {Tannirkulam}, {Webster}, \& {Zhao}}]{2006SPIE.6268E..1PM}
{Monnier}, J.~D., {Pedretti}, E., {Thureau}, N., {et~al.} 2006, in \procspie,
  Vol. 6268, Society of Photo-Optical Instrumentation Engineers (SPIE)
  Conference Series, 62681P

\bibitem[{Monnier {et~al.}(2006)Monnier, Pedretti, Thureau, Berger,
  Millan-Gabet, ten Brummelaar, McAlister, Sturmann, Sturmann, Muirhead,
  Tannirkulam, Webster, \& Zhao}]{PICNIC}
Monnier, J.~D., Pedretti, E., Thureau, N., {et~al.} 2006, Proc.SPIE, 6268, 6268

\bibitem[{Monnier {et~al.}(2012)Monnier, Pedretti, Thureau, Che, Zhao, Baron,
  \& ten Brummelaar}]{MIRC}
Monnier, J.~D., Pedretti, E., Thureau, N., {et~al.} 2012, Proc.SPIE, 8445, 8445

\bibitem[{{Pedretti} {et~al.}(2004){Pedretti}, {Millan-Gabet}, {Monnier},
  {Traub}, {Carleton}, {Berger}, {Lacasse}, {Schloerb}, \&
  {Brewer}}]{2004PASP..116..377P}
{Pedretti}, E., {Millan-Gabet}, R., {Monnier}, J.~D., {et~al.} 2004,
  Publications of the Astronomical Society of the Pacific, 116, 377

\bibitem[{{Scott} {et~al.}(2014){Scott}, {Lhom{\'e}}, {ten Brummelaar},
  {Coud{\'e} du Foresto}, {Millan-Gabet}, {Sturmann}, \&
  {Sturmann}}]{2014SPIE.9146E..1AS}
{Scott}, N.~J., {Lhom{\'e}}, E., {ten Brummelaar}, T.~A., {et~al.} 2014, in
  \procspie, Vol. 9146, Optical and Infrared Interferometry IV, 91461A

\bibitem[{ten Brummelaar {et~al.}(2016)ten Brummelaar, Gies, McAlister,
  Ridgway, Sturmann, Sturmann, Schaefer, Turner, Farrington, Scott, Monnier, \&
  Ireland}]{CHARA}
ten Brummelaar, T.~A., Gies, D.~G., McAlister, H.~A., {et~al.} 2016, Proc.SPIE,
  9907, 9907

\bibitem[{{ten Brummelaar} {et~al.}(2012){ten Brummelaar}, {Sturmann},
  {McAlister}, {Sturmann}, {Turner}, {Farrington}, {Schaefer}, {Goldfinger}, \&
  {Kloppenborg}}]{2012SPIE.8445E..3CT}
{ten Brummelaar}, T.~A., {Sturmann}, J., {McAlister}, H.~A., {et~al.} 2012, in
  \procspie, Vol. 8445, Optical and Infrared Interferometry III, 84453C

\bibitem[{Tsujino {et~al.}(2009)Tsujino, Akiba, \& Sasaki}]{4711091}
Tsujino, K., Akiba, M., \& Sasaki, M. 2009, IEEE Electron Device Letters, 30,
  24

\end{thebibliography}

\begin{appendix}

\section{Gamma distribution}\label{an:gammadist}

\begin{figure}[!ht]
\centering
\includegraphics[width=0.95\columnwidth]{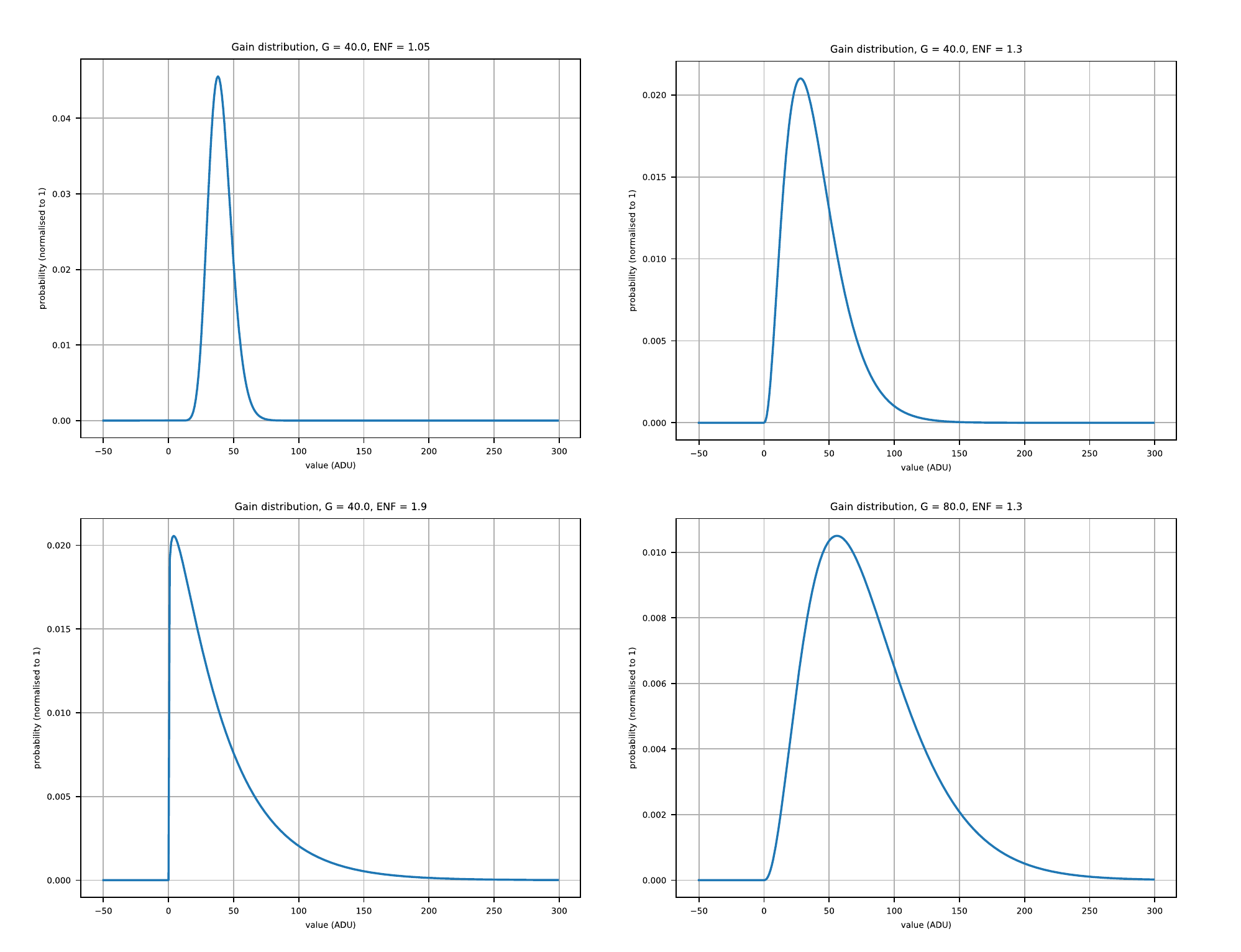}
\caption{Different examples of gain distribution modeled with the Gamma distribution. Upper left corner: $\ENF = 1.05$ and $\Gain = 40$. Upper right corner: $\ENF = 1.3$ and $\Gain = 40$. Lower left corner: $\ENF = 1.9$ and $\Gain = 40$. Lower right corner: $\ENF = 1.3$ and $\Gain = 80$. }\label{fig:Gdist}
\end{figure}

Figure~\ref{fig:Gdist} shows four examples of the Gamma distribution for different values of avalanche gain and ENF. On the upper left corner, we see a Gaussian distribution. On the lower left corner, we are in the regime of an exponential distribution.  The probability density function of the Gamma distribution is defined by
\begin{equation}
    f(x,k;\theta) = \frac{x^{k-1} e^{-x/\theta} }{\theta^k \Gamma(k)}
\end{equation}
for $x > 0$ and $k,\theta > 0$, and where $\Gamma(k)$ is the gamma function. For this distribution, the $k$ parameter describes the shape of the distribution and $\theta$ parameter describes the scale of the distribution.

\section{Different pixels' results}\label{an:diffpix}

Table~\ref{tab:5pixres} summarizes the results obtained for 5 different pixels, in order to quickly assess the homogeneity of the result across the detector.

\begin{table}[!ht]
\caption{Results of gain and ENF for 5 different pixels} 
\label{tab:5pixres}
\begin{center}       
\begin{tabular}{|l|l|l|} 
\hline
\rule[-1ex]{0pt}{3.5ex}  Pixel coordinate & $\Gain / \Gain_{man}$ & \ENF{}  \\
\hline
\rule[-1ex]{0pt}{3.5ex} (67; 7) & 0.72 & 1.49   \\
\hline
\rule[-1ex]{0pt}{3.5ex}  (67; 9) & 0.73 & 1.49 \\
\hline 
\rule[-1ex]{0pt}{3.5ex} (66; 9) & 0.71 & 1.49 \\
\hline
\rule[-1ex]{0pt}{3.5ex}  (67; 11) & 0.69 & 1.44  \\
\hline
\rule[-1ex]{0pt}{3.5ex}   (66; 13) & 0.63 & 1.43  \\
\hline\end{tabular}
\end{center}
\end{table}

The coordinates are measured within the $320 \times 20$ pixel window. The gain factor is computed with the linear fit of the gain of the best-fit model as a function of the manufacturer gain (Fig.~\ref{fig:GGerrbar}). The ENF is obtained with the median of the different ENF measured by the model for various fluxes (Fig.~\ref{fig:ENFerrbar}). The plots shown in the main sections of the paper were for pixel (66; 9). The table~\ref{tab:5pixres} shows that we obtain similar values for other pixels.

\section{Different minimization results}\label{an:minimres}

\begin{figure*}[!ht]
\centering
\includegraphics[scale = 0.9]{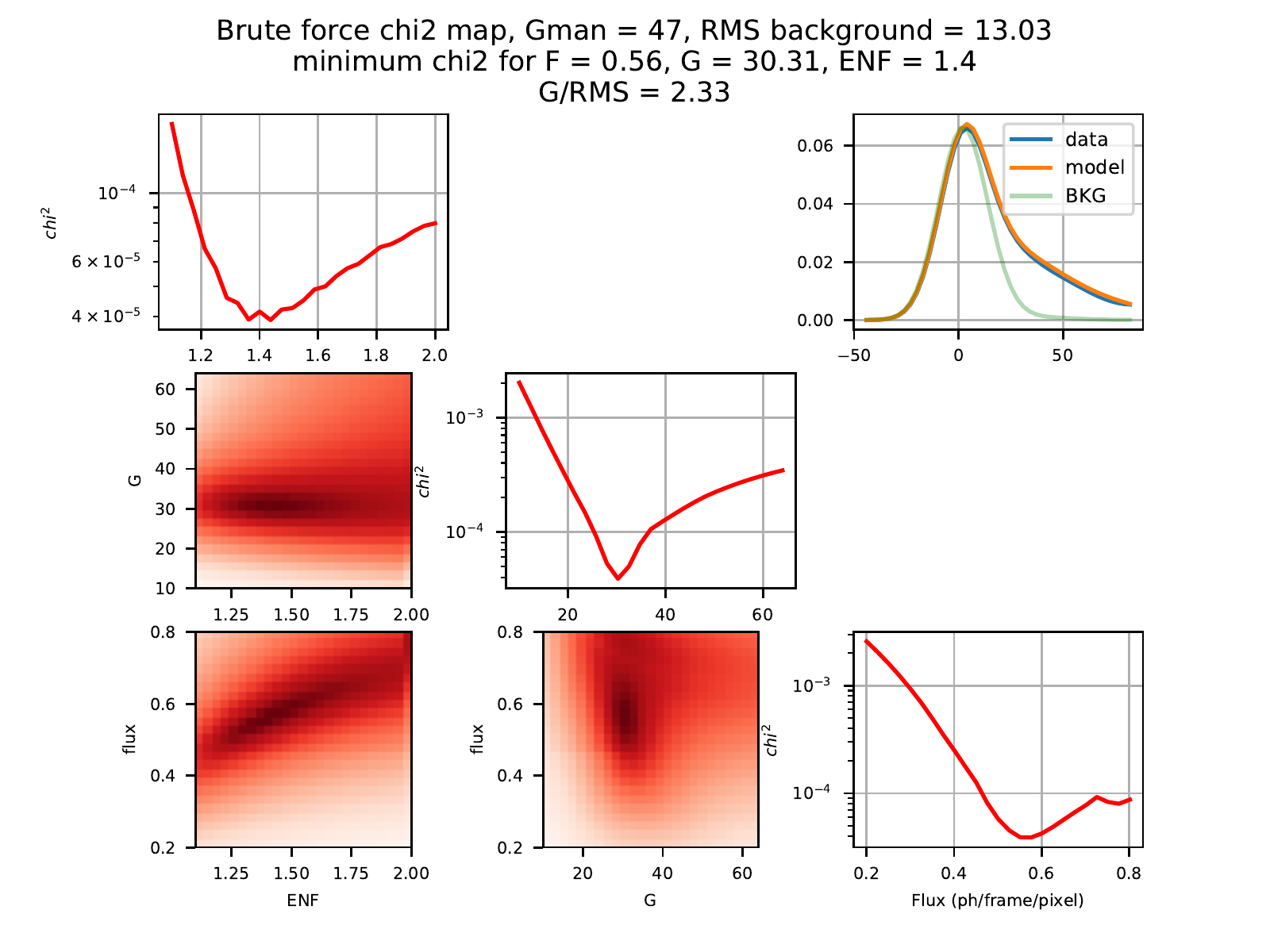}
\caption{\chis{} minimization result example for low flux, low gain}\label{fig:chi2map2-lglf}
\end{figure*}

\begin{figure*}[!ht]
\centering
\includegraphics[scale = 0.9]{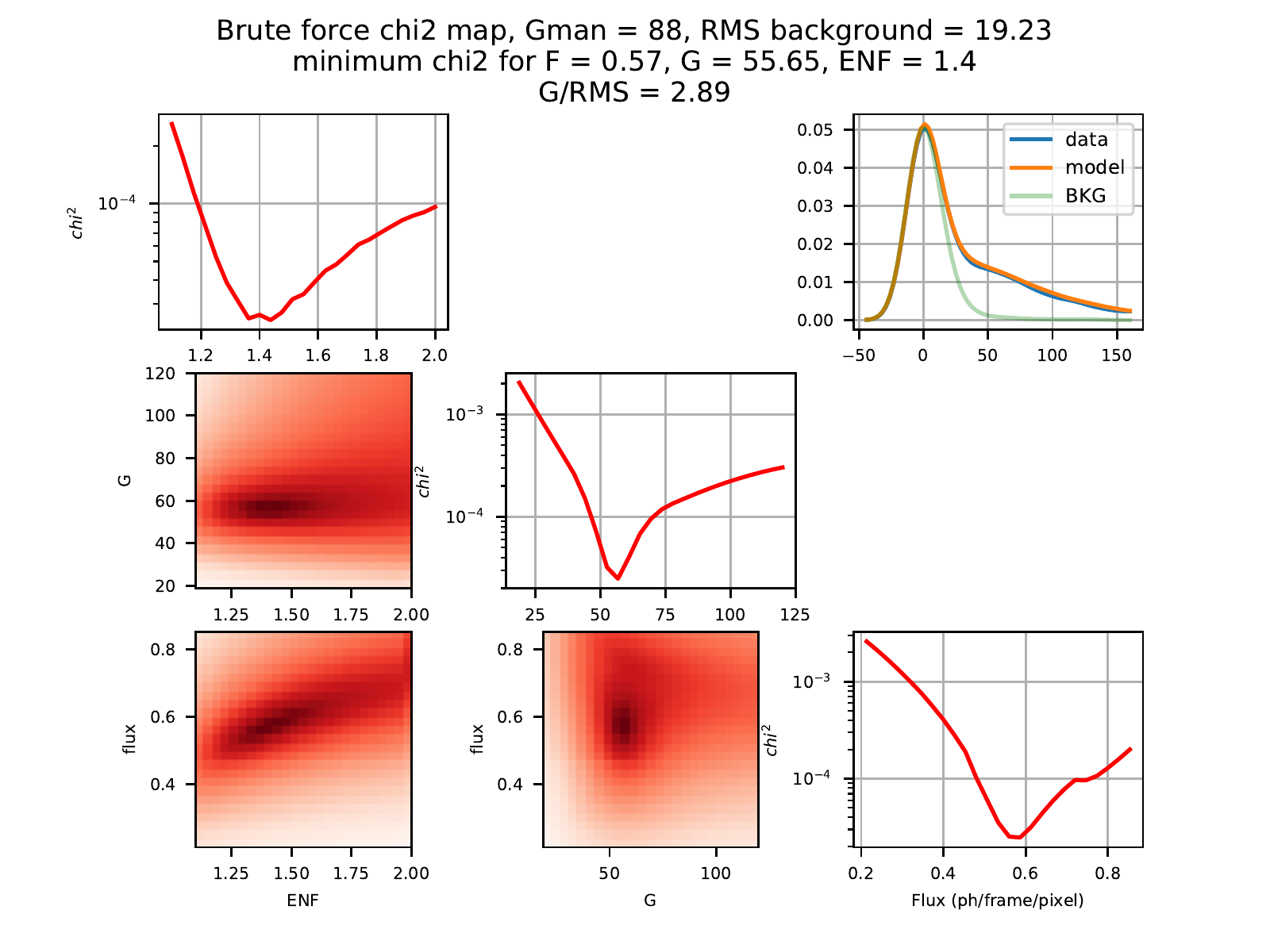}
\caption{\chis{} minimization result example for low flux, high gain}\label{fig:chi2map2-hglf}
\end{figure*}

\begin{figure*}[!ht]
\centering
\includegraphics[scale = 0.9]{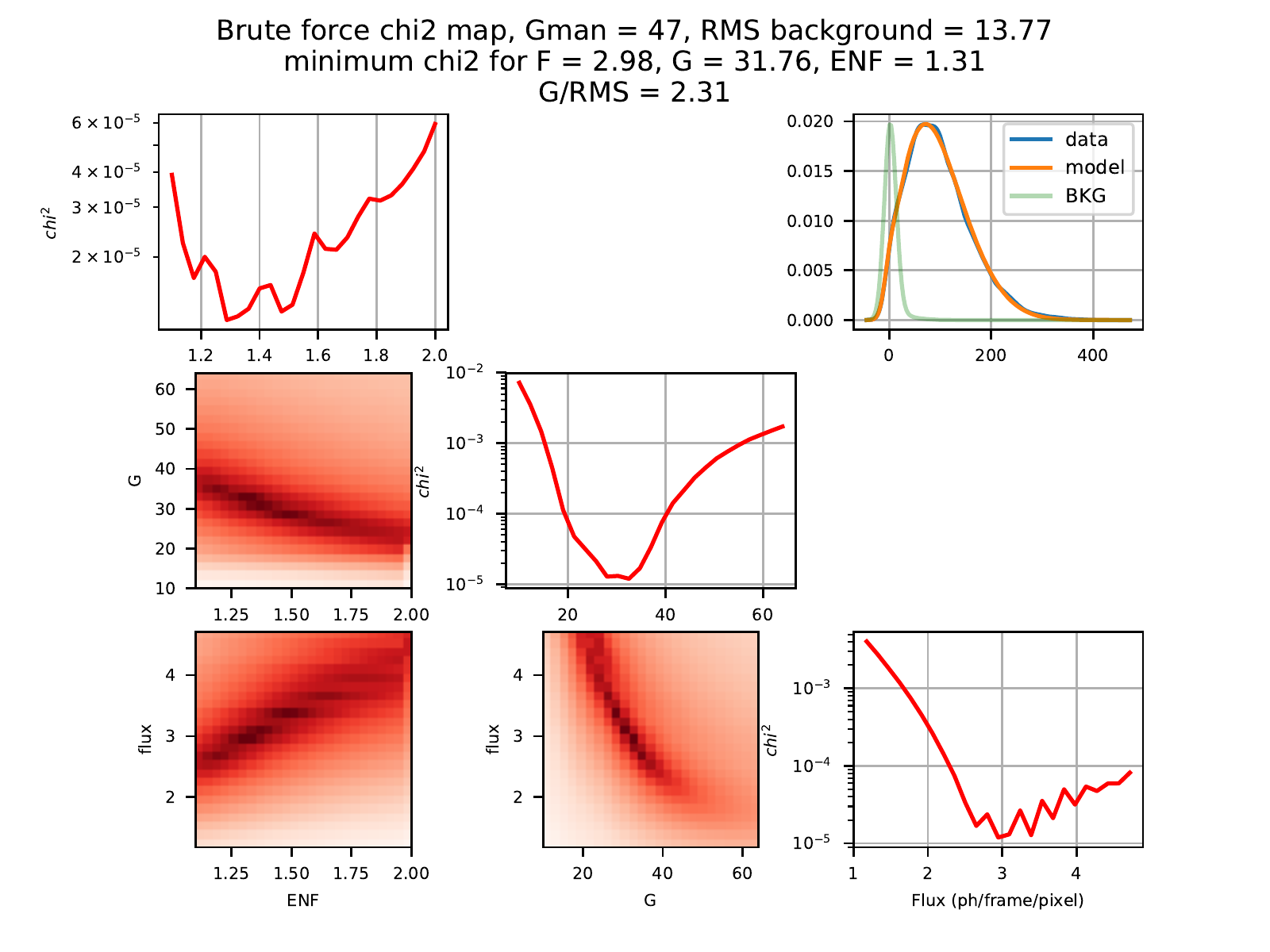}
\caption{\chis{} minimization result example for high flux, low gain}\label{fig:chi2map2-lghf}
\end{figure*}

\begin{figure*}[!ht]
\centering
\includegraphics[scale = 0.9]{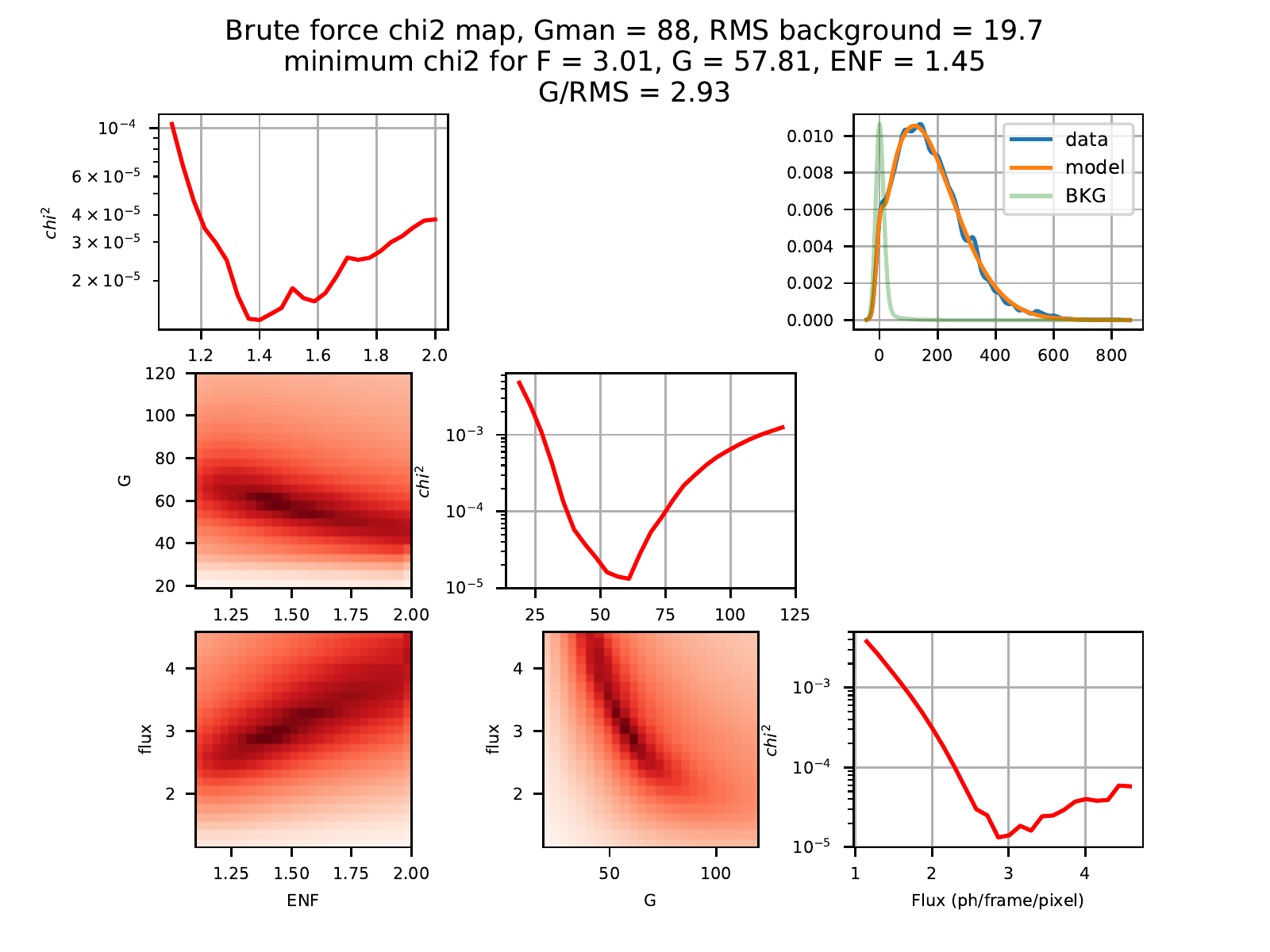}
\caption{\chis{} minimization result example for high flux, high gain}\label{fig:chi2map2-hghf}
\end{figure*}

\begin{figure*}[!ht]
\centering
\includegraphics[scale = 0.9]{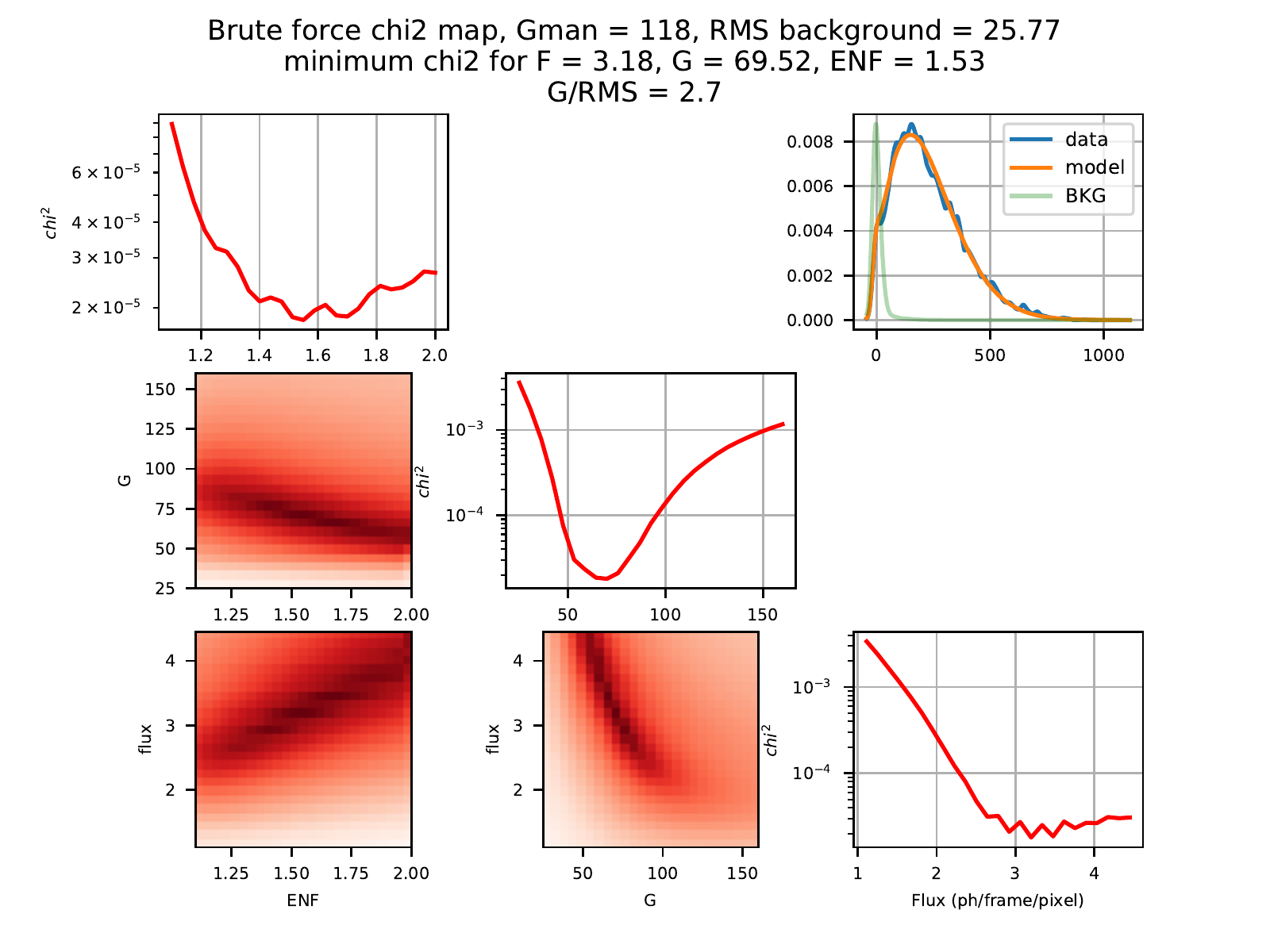}
\caption{\chis{} minimization result example for high flux, highest gain}\label{fig:chi2map2-200}
\end{figure*}

\begin{figure*}[!ht]
\centering
\includegraphics[scale = 0.89]{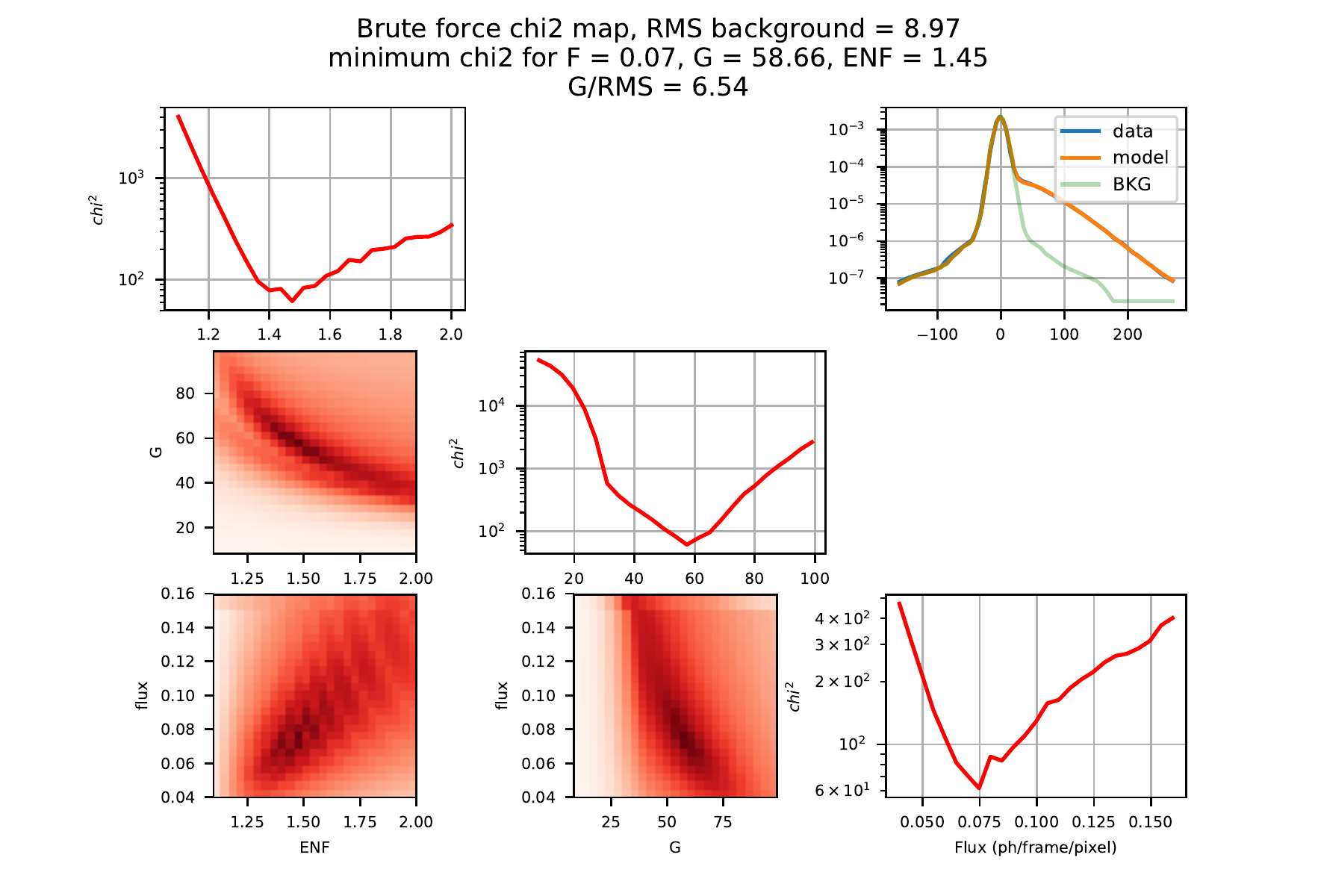}
\caption{\chis{} minimization result for \citet{2018AJ....155..220A} data}\label{fig:chi2map2-atk}
\end{figure*}

This appendix presents the results of minimizations for different typical configurations. Figure~\ref{fig:chi2map2-lglf} is for a low flux and low gain configuration. Figure~\ref{fig:chi2map2-hglf} is for a low-flux/high-gain configuration. Figure~\ref{fig:chi2map2-lghf} is for a high flux and low gain configuration. Figure~\ref{fig:chi2map2-hghf} is for a high-flux/high-gain configuration. Figure~\ref{fig:chi2map2-200} is for a high flux at the maximum gain used for this study. Figure~\ref{fig:chi2map2-atk} is for the data from Fig.\ 6 of \citet{2018AJ....155..220A}.

\end{appendix}

\end{document}